\documentclass[11pt]{article} 

\usepackage[utf8]{inputenc} 

\usepackage[margin=1in]{geometry}
\usepackage{graphicx} 
\graphicspath{ {./imagesColor/} }

\usepackage{booktabs} 
\usepackage{array} 
\usepackage{paralist} 
\usepackage{verbatim} 
\usepackage{subfig} 
\usepackage{xcolor} 
\usepackage{mathrsfs} %
\usepackage{soul} 
\usepackage{verbatim}

\usepackage{fancyhdr} 
\usepackage{sectsty}
\allsectionsfont{\sffamily\mdseries\upshape} 

\usepackage[nottoc,notlof,notlot]{tocbibind} 
\usepackage[titles,subfigure]{tocloft} 

\usepackage{amsmath,amsthm}
\usepackage{amssymb}
\usepackage[hidelinks]{hyperref}

\newtheorem{theorem}{Theorem}
\newtheorem{proposition}{Proposition}
\newtheorem{lemma}{Lemma}
\newtheorem{corollary}{Corollary}
\newtheorem{definition}{Definition}
\newtheorem{conjecture}{Conjecture}

\newtheorem{example}{Example}

\title{\vspace{-1 cm} On the Definition of Black Holes \\ \large Bridging the Gap Between Black Holes and Singularities}
\author{James Wheeler}

\date{} 

\begin{document}

\maketitle

\begin{center}
\vspace{-0.75cm}
Department of Physics, Duke University, Durham, 27710, NC, USA
\\ \vspace{0.25cm}
E-mail: james.c.wheeler@duke.edu
\end{center}

\begin{abstract}
A novel perspective on defining black holes designed to be more broadly applicable outside of asymptotically flat spacetimes, in the context of classical general relativity, is presented, discussed, and characterized. The construction formalizes the heuristic idea that black holes are the ``past Cauchy development of the set of singularities". As such, the formulation depends in a critical way on the identification of singularities, which has been treated in the literature through various boundary constructions. While many of the known boundary constructions (e.g.\@ Schmidt's b-boundary or Geroch's g-boundary) could in principal be cited as underlying one's notion of a black hole, well-known topological concerns leads us to take the perspective that Scott and Szekeres' abstract boundary provides the most natural choice. The framework is utilized to put forward a general, non-IVP formulation of the Weak Cosmic Censorship Conjecture.
\end{abstract}

\section{Introduction}\label{section:intro}

The phenomenon of black holes is among the most surprising and intriguing features of Einstein's General Theory of Relativity-- these causally self-contained regions of spacetime captivate the minds of laypeople and seasoned physicists alike. Among their most widely-discussed features in both of these audiences are the singularities that black holes are generally said to contain. It is a curious situation, then, that the standard definition of what physicists formally mean by the term ``black hole" in classical General Relativity, i.e. the complement of the causal past of an idealized ``future null infinity", neither makes reference to nor is known to guarantee the existence of singularities. It is further curious that while the feature that makes the study of black holes so pertinent is their seemingly generic nature, their widely accepted definition is anything but. Indeed, while cosmological investigations lead us to believe the universe at large is definitively not asymptotically flat given the presence of dark energy (\cite{aghanim2020planck}, \cite{riess1998observational}, \cite{perlmutter1999measurements}) it continues to be the case that the most standard and robust classical definition of black holes hinges on asymptotic flatness. Nevertheless, this standard approach and existing variations upon it have proved remarkably fruitful and insightful, and so we by no means wish to supplant them: rather, it is the objective of this paper to provide an additional and complementary perspective on defining black holes which ameliorates the above curiosities.

Though both black holes and singularities in general relativity have a storied history, the programs of research devoted to them diverged somewhat following the seminal result associating them, Penrose's widely celebrated Incompleteness Theorem \cite{penrosesingularity}, for which he was partially awarded the 2020 Nobel Prize in Physics. The phenomenon of singularities in general relativity has been investigated in its own right, independently of any black holes to which they may be associated, for over half a century through the development of various so-called ``boundary constructions". This course of study seemingly began around 1960 with Szekeres' {\it On the Singularities of a Riemannian Manifold} \cite{szekeres}, which has been succeeded by various constructions seeking to associate to a general spacetime manifold $(M,g)$ a topological space $\overline{M}$ consisting of $M$ together with a collection of boundary points $\partial M$, some or all of which are identified as singularities. Among the best-known and most influential such constructions were Geroch's geodesic boundary (g-boundary)  \cite{gerochboundary}, Schmidt's bundle boundary (b-boundary) \cite{schmidtbb}, and Scott and Szekeres' abstract boundary (a-boundary) \cite{scottszekeres}. 

While the various boundary constructions get rather technically involved, each achieves a somewhat simple goal: by constructing an appropriately physical topology on $\overline{M}$, they identify what it means for a point in the spacetime manifold $M$ to be ``near" a singularity. Crucially, they do this in a manner that is entirely intrinsic to $M$-- no additional structure on the spacetime manifold need be invoked in order to make the identification. In this sense, programs for identifying ``where" singularities lie in General Relativity are available for completely general spacetimes. This will be a key ingredient in our definition of a black hole for a general spacetime.

Meanwhile, the study of black holes and their features has been principally treated, or at least motivated, via their original characterization, due to Penrose \cite{penroseconformalinfinity}, as that subset of spacetime which cannot be seen from infinity, provided one restricts spacetime to be of a form in which one has a particularly well-behaved notion of infinity. This is discussed at length in classic texts \cite{hellis, waldgr} and reviewed below. This has been plenty sufficient for many observational and numerical modeling investigations, as well as theoretically useful to proving many important and well-known properties of black holes, e.g.\@ thermodynamic properties \cite{bardeen1973four}, in particular Hawking's area theorem \cite{hawking1972black}, and the Riemannian Penrose inequality \cite{bray2001proof, huisken2001inverse}. Beyond the aesthetic appeal of having a consistent definition which makes unequivocal sense in the context of cosmology and beyond, however, it is the perspective of the author that it is of great theoretical interest to ask how such properties may or may not extend to a broader description of black holes, as well as to investigate what insight might be gained into significant open questions when formulated under such a description. 

Indeed, we use our definitions to put forward a more comprehensive formulation of the Weak Cosmic Censorship Conjecture in Section \ref{section:weakcosmiccensorship}, a problem to which our perspective on black holes is uniquely suited due to its intimate ties to singularities. We argue there that such a reformulation is needed to handle naked singularities that cannot be described via the standard IVP formulation. Such naked singularities include those present in the Vaidya spacetimes, which we have recently demonstrated \cite{wheelervaidya} to be significantly more generic than previously discussed in the literature.

The problem of defining black holes is one of many facets, of course: many different formal characterizations have been put forward in different subfields of physics, and even more numerous heuristic characterizations are used in practice. For a discussion of the history of this topic in the various subfields of physics, see Curiel's essay \cite{curiel}. This paper works strictly within the domain of classical general relativity, wherein alternatives to the standard paradigm are generally based upon local convergence properties of geodesics and suggesting one identify black holes by a new type of horizon loosely modeled after apparent horizons (prototypes were provided by Krolak \cite{krolak1982definitions}, Hayward \cite{hayward1994general, hayward2000black}, and Ashtekar et al.\@ \cite{ashtekar2000generic}). These are generally conceived with practical numerical and semi-classical thermodynamic concerns in mind. While we comment on these very briefly in Section \ref{section:standard}, see \cite{ashtekar2004isolated, nielsen2009black} for reviews of the status of such horizon-based approaches.

This paper will be organized as follows. In Section \ref{section:standard}, we give a brief review of the standard characterization of black holes (following Hawking and Ellis \cite{hellis}). In Section \ref{section:blackhole}, we motivate and put forward the key definitions of interest, given that one has a notion of ``where" singularities are (e.g.\@ via a boundary construction as discussed above), leading to the proposed definition of black holes. We illustrate these definitions in several examples highlighting both similarities and differences from the classic definition. In Section \ref{section:singularnbhd}, we briefly weigh the merits of various boundary constructions before committing to the a-boundary (reviewed in the appendix) and using it to formalize singular neighborhoods. In Section \ref{section:results}, various results surrounding the proposed notion of black hole are presented, proven, and discussed. In Section \ref{section:weakcosmiccensorship}, we discuss application of the formalism to the Weak Cosmic Censorship Conjecture.

\section{The Classic Perspective}\label{section:standard}

The classic formalization of the notion of a black hole depends on a hierarchy of technical mathematical structure building up to formalizing {\it asymptotic flatness} of the spacetime $(M,g)$, meant to encode that the spacetime is comprised of an isolated or bounded system of matter. The precise details of this hierarchy are subject to some variation (see, for example, \cite{penroseconformalinfinity} versus \cite{hellis} versus \cite{waldgr}, or \cite{dafermos2013lectures} for a more modern perspective), but the broad strokes are essentially the same. We present the hierarchy as described in Hawking and Ellis \cite{hellis}, so definitions in this section are taken from there. At the base is an {\it asymptotically empty and simple} spacetime:

\begin{definition}\label{definition:asymptoticallysimple}
A time- and space-orientable spacetime $(\widetilde M,\tilde g)$ is \ul{asymptotically empty and simple} if there exists a strongly causal spacetime $(\widehat{M}, \hat g)$ and a conformal embedding $\phi: \widetilde M \to \widehat M$, under which $\overline{\phi(\widetilde{M})}$ is a submanifold with smooth boundary $\partial \phi(\widetilde M) $, satisfying:
\begin{enumerate}[(1)]
\item There is a sufficiently smooth function $\Omega : \widehat M \to \mathbb{R}$ satisfying $\Omega > 0$ on $\phi(\widetilde M)$ such that $(\Omega \circ \phi)^2 \tilde g = \phi^* \hat g$.
\item $\Omega = 0$ and $d \Omega \neq 0$ on $\partial \phi (\widetilde M) $.
\item Every inextendible null geodesic in $\widetilde M$ has two endpoints on $\partial \phi (\widetilde M)$.
\item There exists an open set $\widehat U \subset \widehat M$ containing $\partial \phi (\widetilde M)$ such that $\widetilde{\text{Ric}} = 0$ on $\widetilde U := \phi^{-1}(\widehat U)$.
\end{enumerate}
\end{definition}

Condition (1) indicates that $(\widehat M, \hat g)$ encodes the global causal structure of $(\widetilde M, \tilde g)$; condition (2) ensures that $\partial \phi(\widetilde M)$ is indeed at ``infinity" with respect to $\widetilde M$ in the sense that any null geodesics in $\widetilde M$ approaching $\partial \phi(\widetilde M)$ must have infinite affine parameter; condition (3) ensures that $\partial \phi(\widetilde M)$ includes the entirety of infinity; condition (4) ensures, given the Einstein equation, that the matter under consideration is bounded away from infinity. For more discussion of these conditions and their motivations and implications, see \cite{hellis}. Perhaps most importantly, they imply that $\partial \phi(\widetilde M)$ is a null hypersurface in $\widehat M$ comprised of two disconnected pieces, {\it future null infinity} $\mathscr J^+$ and {\it past null infinity} $\mathscr J^-$, the collections of future and past (respectively) endpoints in $\widehat M$ of inextendible null geodesics in $\widetilde M$. It was among the first successes of boundary constructions that they demonstrated that the closure of $\phi(\widetilde M)$ in $\widehat M$ is unique, e.g.\@ independent of the choice of $\phi$ and $\widehat M$ (\cite{gerochboundary}, \cite{schmidt1991uniqueness}), so that the structure at infinity really is intrinsically meaningful to $\widetilde M$ in this setting.

While effective at capturing many of the desired qualities, this definition is apparently too restrictive in that it manifestly requires $\widetilde M$ to be null geodesically complete, and so cannot be applied directly to the standard black hole solutions (e.g Schwarzschild, Reissner-Nordstrom, Kerr), or any spacetimes including trapped surfaces and satisfying the hypotheses of Penrose's Incompleteness theorem. The standard remedy is to put forward the notion of a {\it weakly asymptotically simple and empty} spacetime:

\begin{definition}\label{definition:weaklyasymptoticallysimple}
A spacetime $(M,g)$ is said to be  \ul{weakly asymptotically empty and simple} if there is an open set $U \subset M$ and an asymptotically empty and simple spacetime $(\widetilde M, \tilde g)$ with an open neighborhood $\widehat U$ of $\partial \phi(\widetilde M)$ in $\widehat M$ such that $(U,g)$ is isometric to $(\widetilde U, \tilde g)$, where $\widetilde U : = \phi^{-1}(\widehat U)$.
\end{definition}

This definition simply describes a spacetime which looks weakly asymptotically empty and simple in some region including an apparently full notion of infinity. Any choice of such a $U \subset M$ and associated asymptotically empty and simple spacetime $(\widetilde M, \tilde g)$ endows $M$ with a future null infinity $\mathscr{J}^+ \subset \widehat M$, whose causal relation to $M$ is unambiguous in that we can make sense of the causal past of $\mathscr{J}^+$ in $M$. To be painfully explicit, denoting by $\psi : U \to \widetilde U$ the isometry referred to in the definition, this relation is simply
$$J^-(\mathscr{J}^+) := J^- \left( (\phi \circ \psi)^{-1} \left( \widehat J^-(\mathscr{J}^+) \right) \right),$$
where $\widehat J^-(\cdot)$ refers to taking the causal past in $\widehat M$. There {\it is} ambiguity, however, implicit in the choice of $U$ and $(\widetilde M, \tilde g)$ above. To make sense of $J^-(\mathscr{J}^+)$, then, one must fix this choice.

At this point, the core structures are in place to allow one to define a black hole in the classic sense. It is simply that subset of a weakly asymptotically empty and simple spacetime which cannot ``escape to infinity". That is,

\begin{definition}\label{definition:classicblackhole}
In a weakly asymptotically empty and simple spacetime $(M,g)$, the (classic) \ul{black hole region} with respect to a future null infinity $\mathscr{J}^+$ is defined as 
$$\mathscr{B}_c := M \backslash J^-(\mathscr{J}^+).$$
\end{definition}

The term ``black hole" now typically refers to the intersection of $\mathscr{B}_c$ with a spacelike hypersurface $\Sigma \subset M$. Though it is perfectly meaningful and satisfies the physical criteria of what a black hole should be, this definition is somewhat pre-emptive in that one typically restricts a bit more the class of spacetimes within which one identifies $\mathscr{B}_c$ so as to render provable the various celebrated results on black hole properties and dynamics. In particular, most such results require working within the following domain:

\begin{definition}\label{definition:asymptoticallypredictable}
A weakly asymptotically empty and simple spacetime $(M,g)$ is said to be 
\ul{future asymptotically predictable} provided that there exists an edgeless acausal subset $\mathscr{S} \subset M$ such that $\mathscr{J}^+$ is in the closure of $D^+(\mathscr{S})$, the future Cauchy development of $\mathscr{S}$, in $\widehat M$.
\end{definition}

This condition essentially encodes that there is a three-dimensional spacelike hypersurface in $M$ from which $\mathscr{J}^+$ can be evolved. This is interpreted as meaning that there are no non-initial ``naked singularities", singularities visible from $\mathscr{J}^+$ to the future of this hypersurface. This condition is generally considered reasonable, then, since it is seen as tantamount to assuming the veracity of the Weak Cosmic Censorship Conjecture. With it (or perhaps with a slight strengthening), one may prove \cite{hellis} that closed trapped surfaces, outer trapped surfaces, and apparent horizons in $D^+(\mathscr{S})$ always lie in $\mathscr{B}_c$, and that black hole boundaries increase in area over time. This is then the basis upon which many take trapped surfaces and variations on the concept (marginally trapped surfaces, apparent horizons, minimal surfaces in time-symmetric initial data, etc.) to be quasi-local stand-ins for the concept of a black hole when working in particular contexts, e.g.\@ numerical relativity or initial data formulations of the Penrose inequality. Indeed, the extant attempts at generalizing the term ``black hole" outside of asymptotically flat contexts (\cite{krolak1982definitions},\cite{hayward1994general},\cite{ashtekar2004isolated},\cite{bengtsson2011region}) are built around these ideas.

The most obvious limitation of Definition \ref{definition:classicblackhole} is its dependence on the hierarchy of structure built up over Definitions \ref{definition:asymptoticallysimple}, \ref{definition:weaklyasymptoticallysimple}, and \ref{definition:asymptoticallypredictable}. These are significant constraints on $(M,g)$ invoking considerable extrinsic structure which, while inarguably useful for investigations both numerical and theoretical into the properties of approximately isolated systems in general relativity, seem well beyond what should be required to characterize the intuitive concept of a black hole as a region of no escape, and are manifestly inapplicable to cosmological situations which cannot be well-approximated as isolated (e.g.\@ primordial black hole formation). We take the perspective that this should be characterizable in a general spacetime in a completely intrinsic manner. There is also implicit ambiguity in Definition $\ref{definition:classicblackhole}$ inherited from that in the notion of $\mathscr{J}^+$ induced by Definition \ref{definition:weaklyasymptoticallysimple}, in that there may well be multiple distinct neighborhoods $U \subset M$ isometric to a neighborhood of infinity in some $\widetilde M$. This is not at all unfamiliar, as it is an immediate feature of the maximal extensions of the most standard black hole spacetimes that there are multiple distinct future null infinities, and so associated to each of them there is a distinct black hole region (though they may partially overlap, as in the maximally extended Schwarzschild case). ``The" classic black hole region $\mathscr{B}_c$ of a given spacetime is then a matter of perspective.

While, as previously mentioned, some more broad approaches to defining black holes have been put forward, this broadening has generally come at the cost of losing direct association to a core feature of the intuitive physical description: alternative approaches are largely based on using local convergence properties of geodesics to identify new types of horizons (\cite{ashtekar2000generic, hayward1994general, hayward2000black, krolak1982definitions}), which may not in general enjoy the ``trapping" features with respect to infinity guaranteed in the future asymptotically predictable setting. Sets identified via these conditions simply need not bound a region that is meaningfully ``small", as one might like. Indeed, the trapped surfaces in the maximally extended Kerr spacetime have the property that their causal future always includes some $\mathscr{J}^+$ (infinitely many, in fact-- see Figure \ref{fig:kerr}), so it is certainly not the case that trapped surfaces bound a region that is ``trapped" in any universal sense. Without the structure of asymptotic flatness to pick out an infinity with respect to which they are ``trapped", then, these surfaces and their generalizations lose their direct link to a significant component of the black hole concept. This is further exemplified in non-singular constructions conforming to some such definitions (but not Definition \ref{definition:classicblackhole}), such as Hayward's \cite{hayward2006formation}, wherein every timelike curve exits the region identified as a black hole. Depending on one's objective, this may not be so dire a cost, particularly from a semi-classical perspective seeking to accommodate escaping Hawking radiation (as in Hayward's case). Moreover, these convergence characterizations have certainly been valuable to numerically identifying black holes, investigating primordial black holes in practice \cite{harada2013threshold}, and deciding how to best formulate black hole thermodynamics. There are contexts within which they fall short, however -- they are not suited for purely classical questions such as weak cosmic censorship, for example--, and so we feel something is still missing in the effort to fully, intuitively, and rigorously characterize black holes in a general setting.

\section{What Makes a Black Hole?} \label{section:blackhole}

\subsection{A Dual Perspective}

Having seen the standard construction which we hope to build upon, we should now step back and reflect: what is the essence of a black hole which it captured, and so which any extension must capture as well? The heuristic layman's definition, of course, is that a black hole is {\it a region of spacetime in which gravity is so strong that light cannot escape}-- this is perhaps the most crucial and recognizable feature of a black hole. By itself, though, this is clearly not sufficient as a technical definition, as it is true directly by definition of any future set whatsoever (a set $F \subset M$ such that $J^+(F) = F$) that light cannot escape from it. Such sets include $J^+(p)$ for any $p \in M$, $M \backslash J^-(p)$ for any $p \in M$, or even the entirety of $M$ itself; these clearly should not be called black holes in general. There is another feature of a black hole which a complete definition should capture, however: a black hole is ``small" in some appropriate sense. Taking all of $M$ most glaringly demonstrates the need for this restriction, as it is not so insightful to note that light cannot escape the whole of spacetime. This is only a meaningful constraint when the light is somehow bounded.

How can we mathematically capture the idea of a black hole's boundedness? The standard construction's answer to this question is to appeal to a notion of infinity: the light of a standard black hole is ``bounded" in the sense that it cannot reach the infinity provided by the structure of asymptotic flatness. We hope to find an answer to this question even when asymptotic flatness cannot provide such an infinity. Extant attempts are content with a local stand-in for boundedness through geodesic convergence, while we seek to preserve the global outlook. To do so, we shall take a perspective somewhat dual to the standard one-- we ask not which light cannot reach infinity, but which light {\it must} reach a singularity. Indeed, if light approaching the edge of spacetime is not to reach any notion of infinity, then a singularity seems the only alternative.

\begin{figure}[t]
\centering
\includegraphics[width=10cm]{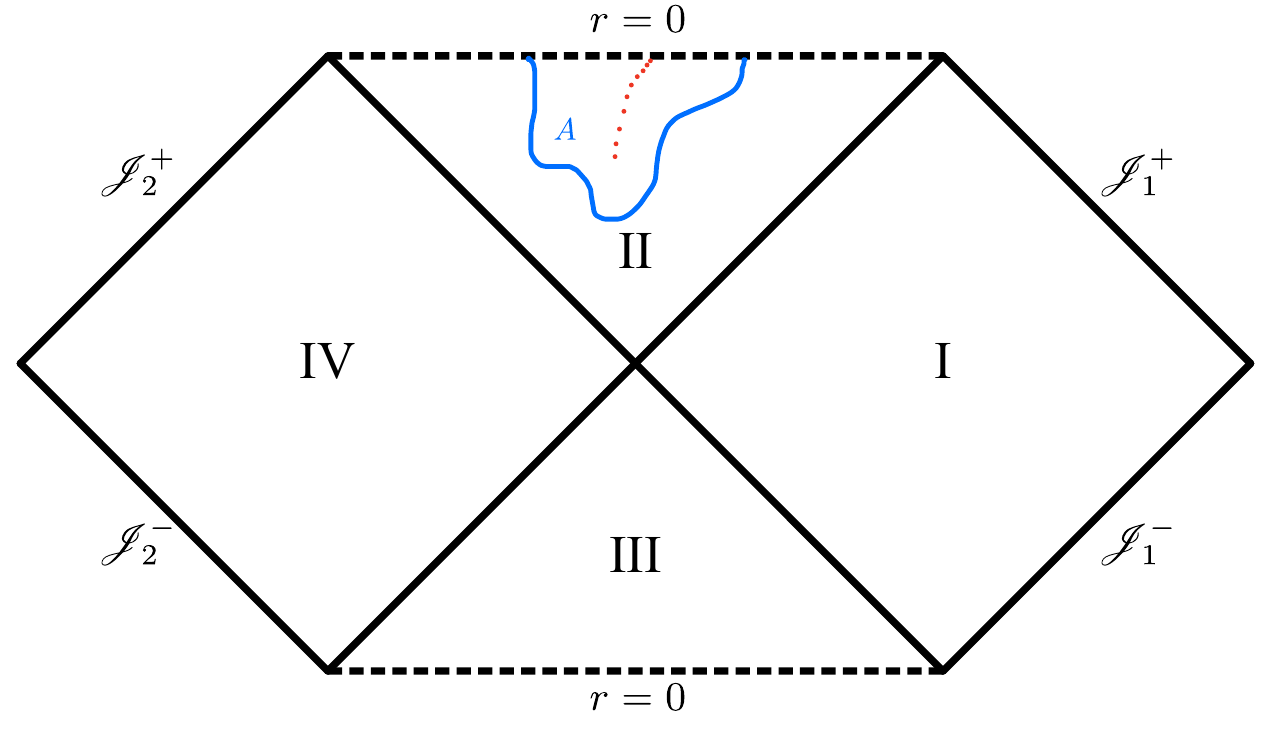}
\caption{\footnotesize Schwarzschild Penrose diagram depicting a sequence in the black hole, region II, with no convergent subsequence. Also shown is a closed set $A$ in region II containing the sequence which should be ``small" in whatever sense the black hole is.}
\label{fig:noncompact}
\end{figure}

The topologist's favorite characterization of a ``small" or ``bounded" set is, of course, compactness. In the context of manifolds, which are topologically sequential, compact sets are simply those in which every sequence of points has a convergent subsequence. This nicely describes that a compact set is ``small" in the sense that an infinite collection of points cannot spread out-- they must tend to gather around at least one point in order to all fit in the set. This would be a fine characterization of ``small", except that it utterly fails to capture the smallness of black holes. Indeed, in the prototypical case of the Schwarzschild spacetime, one may take any sequence of points in the black hole region along which $r \to 0$ monotonically, and this can have no convergent subsequence. This is demonstrated in the Penrose diagram of the Schwarzschild spacetime in Figure \ref{fig:noncompact}: while the sequence depicted in red apparently converges to a point on the $r = 0$ line of singularity, this point is not in the spacetime, so the sequence has no subsequence which converges to a point in the spacetime, and hence the closed set $A$ which contains the sequence is non-compact. Being a subset of the black hole, however, $A$ must be ``small" in whatever sense the black hole is.

This apparent failure of compactness can be remedied. What we notice is that the failure of a sequence in the depicted set $A$ to have a convergent subsequence can {\it only} happen in this way-- problematic sequences {\it must} approach the $r=0$ singularity. So that we might convey the core content and pictorial intuition of our construction undistracted by technicalities, let us for the moment sweep under the rug the problem of identifying ``where" the singularities are, to be returned to in Section \ref{section:singularnbhd}. That is, let us assume that we are gifted the collection $\mathscr{U}$ of open subsets of $M$ which are deemed to envelop all singularities of $M$, in the sense that any sequence in $M$ which ``approaches a singularity", whatever that might mean, must eventually enter and remain inside each $U \in \mathscr{U}$. These open sets will be called {\it singular neighborhoods}. 

\begin{figure}[t]
\centering
\includegraphics[width=10cm]{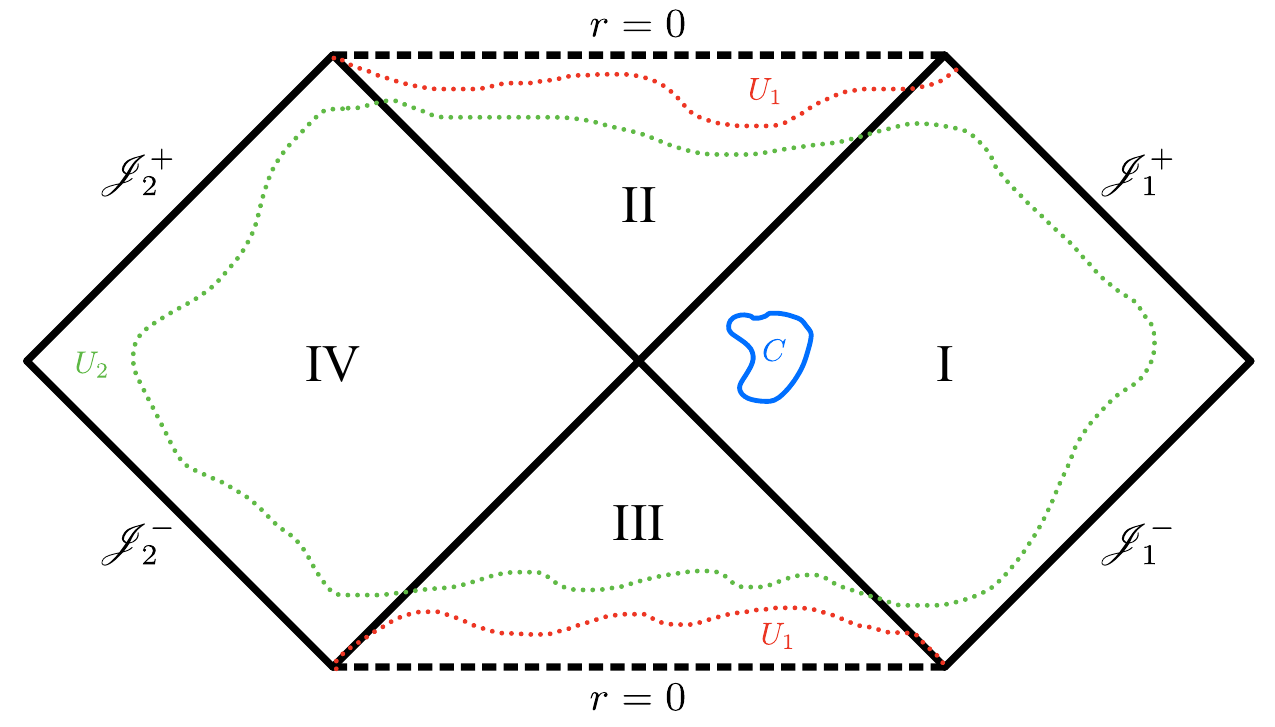}
\caption{\footnotesize Examples of singular neighborhoods in the Schwarzschild spacetime. $U_1$ is the region above the top and below the bottom red dotted curves. $U_2$ is the exterior of the closed green dotted curve. The complement of the compact set $C$ is also a singular neighborhood. All of these ``surround" the singular $r = 0$ lines by as much as possible from within $M$. If one removes any of these from the set $A$ of Figure \ref{fig:noncompact}, it becomes compact.}
\label{fig:singularnbhd}
\end{figure}

As some examples, the family $\mathscr{U}$ of singular neighborhoods should always include the entire manifold, i.e. $M \in \mathscr{U}$, as well as the complement of any compact set. It is closed under union with arbitrary open sets and finite intersection. In a ``nonsingular" (again, whatever that might mean-- though it should certainly include Minkowski space) spacetime, $\mathscr{U}$ should simply be the entire topology on $M$, as the condition of approaching a singularity is then null. Some examples for the Schwarzschild spacetime are depicted in Figure \ref{fig:singularnbhd}.

Armed with the collection of singular neighborhoods, we are now prepared to introduce our means of characterizing the ``smallness" of the set $A$ depicted in Figure \ref{fig:noncompact}:

\begin{definition} \label{definition:scomp}
Let (M,g) be a spacetime manifold. A closed set $A \subset M$ is called \ul{singularly compact} if $A\backslash U$ is compact for every singular neighborhood $U \in \mathscr{U}$.
\end{definition}

This definition is saying, then, that singularly compact sets are precisely those that are compact outside of singular behavior, i.e. except for the possibility of sequences approaching singularities. Indeed, it is immediate from the definition that a closed set $A$ is singularly compact iff every sequence in $A$ with no convergent subsequence eventually enters and remains in every singular neighborhood. The family of singularly compact sets is closed under intersections and finite unions, and closed subsets of singularly compact sets are singularly compact; compact sets are trivially singularly compact. In a ``nonsingular" spacetime, the singularly compact sets are precisely the compact sets. Physically, singularly compact sets are simply ``finite" or ``small" sets that may be arbitrarily close to any singularities.

\begin{definition}\label{definition:blackregion}
Let $\mathscr{F}$ be the family of singularly compact future sets, i.e. singularly compact sets $A \subset M$ satisfying $J^+(A) = A$. The \ul{black region} $\mathscr{B} \subset M$ is given by 
$$\mathscr{B} := \bigcup_{A \in \mathscr{F}} A.$$
A connected component of the intersection of $\mathscr{B}$ with a spacelike hypersurface is called a \ul{black hole}.
\end{definition}

This definition is the proposed formal characterization of black holes, applicable to any spacetime, which we wish to put forward. It stipulates that a point $p \in M$ is in $\mathscr{B}$ if and only if it lies in a singularly compact future set, a ``small" set from which no causal signal can escape. One could, of course, define a white region $\mathscr{W}$ and its associated white holes in a completely analogous manner by replacing future sets with past sets everywhere, but we will not treat this explicitly. Notice that this definition makes no reference to a particular choice of region at infinity, as is implicit in Definition \ref{definition:classicblackhole} via its invocation of $\mathscr{J}^+$, so these two definitions are certainly not identifying the same subset of $M$ (when Definition \ref{definition:classicblackhole} is applicable). Instead, ours hopes to identify those points which would be said to be in a black hole with respect to {\it any} portion of infinity, which any observer should agree would be in a black hole. We explore this distinction further in the examples to follow. First, a simple consequence of the definition:

\begin{lemma}\label{lemma:blackcausalfuture}
For any $p \in M$, $p \in \mathscr{B}$ iff $\overline{J^+(p)}$ is singularly compact.
\end{lemma}

\proof If $p \in \mathscr{B}$, $p \in A$ for some singularly compact set $A$ satisfying $J^+(A) = A$, and hence $J^+(p) \subset A$. Since $A$ is closed, $\overline{J^+(p)} \subset A$, so for every singular neighborhood $U$, $\overline{J^+(p)} \backslash U$ is a closed subset of the compact set $A \backslash U$, hence is compact, showing $\overline{J^+(p)}$ is singularly compact.

Conversely, if $\overline{J^+(p)}$ is singularly compact, then since $J^+(\overline{J^+(p)}) = \overline{J^+(p)}$, we have $\overline{J^+(p)} \in \mathscr{F}$, so $p \in \overline{J^+(p)} \subset \mathscr{B}$.

\qed

This straightforward lemma indicates that a point lies in $\mathscr{B}$ iff its causal future is ``small" in the sense of singular compactness, so that the question of a point's being in the black region is a question about its causal future. This means that the characterization for a point $p \in M$ is somewhat localizable, in that if one modifies the spacetime only outside if $\overline{J^+(p)}$, it should not change whether $p$ is deemed to be in $\mathscr{B}$. If one perturbs or adjusts the Schwarzschild spacetime in the $r>2m$ region in any way (even destroying asymptotic flatness), for example, the black piece of the $r<2m$ region should still be deemed to be in a black hole. Precise results of this nature can only be formally proven, of course, once one has precisely described singular neighborhoods. We detail one way of doing so in Section \ref{section:singularnbhd}, but first we explore the intuition and heuristics of Definition $\ref{definition:blackregion}$ in some standard spacetimes.

\subsection{Examples}
\begin{figure}[t!]
\centering
\includegraphics[width=10cm]{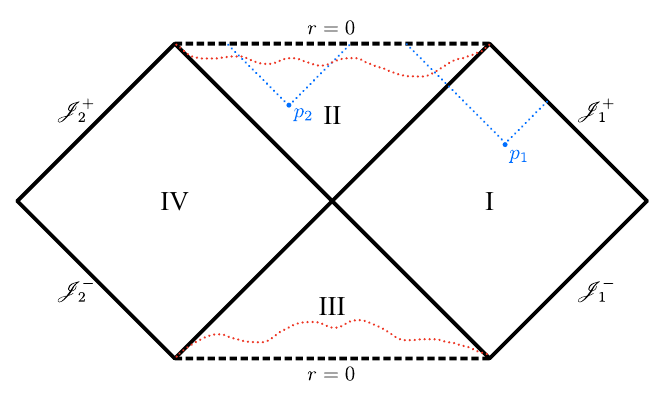}
\caption{\footnotesize Identifying the black region in the Schwarzschild spacetime. The red dotted curves form the boundary of a typical singular neighborhood $U$ (similar to $U_1$ in the previous figure), while the blue dotted lines are the boundaries to the causal futures of the points $p_1$ and $p_2$. These causal futures are then everything above the boundaries.}
\label{fig:schwarzschildEx}
\end{figure}

\begin{figure}[b!]
\centering
\includegraphics[width=9cm]{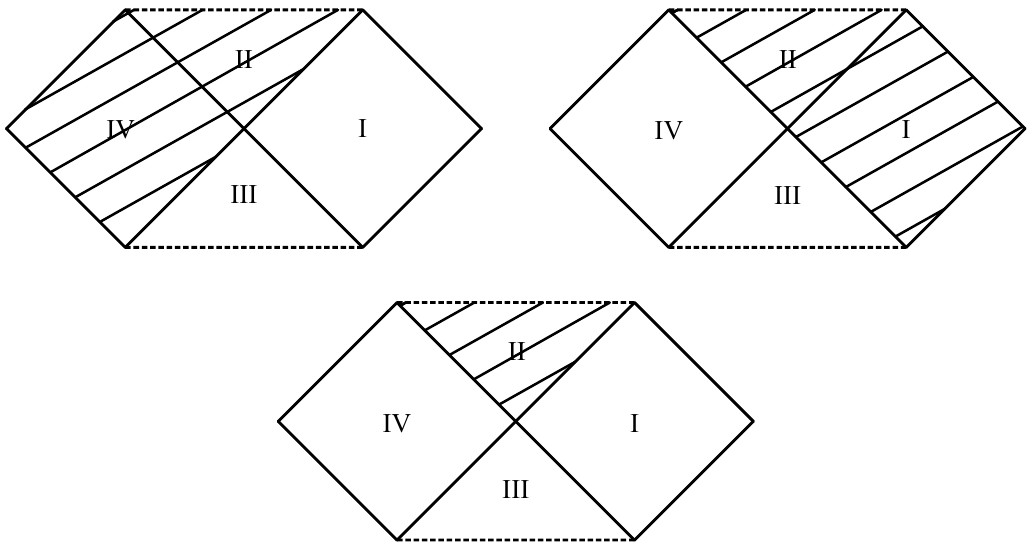}
\caption{\footnotesize The two possible classic black hole regions $\mathscr{B}_c$, shown shaded, in the Schwarzschild spacetime (top), contrasted with the single black region $\mathscr{B}$ of Definition \ref{definition:blackregion} (bottom).}
\label{fig:schwarzschildEx2}
\end{figure}

\begin{figure}[t!]
\centering
\includegraphics[width=15cm]{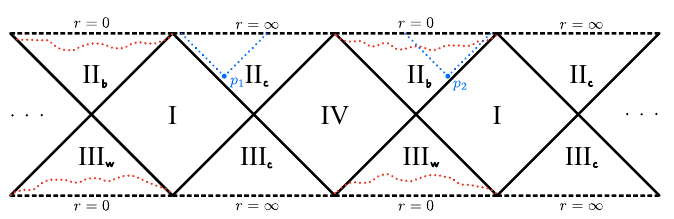}
\caption{\footnotesize Identifying the black region in the deSitter Schwarzschild spacetime.}
\label{fig:deSitterSchwarzschild}
\end{figure}

\begin{figure}[b!]
\centering
\includegraphics[width=16cm]{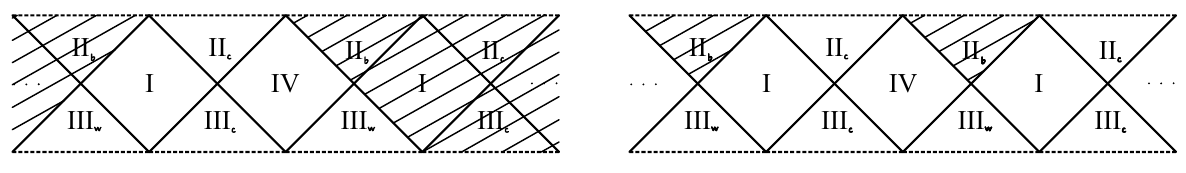}
\caption{\footnotesize A classic black hole region $\mathscr{B}_c$ in the deSitter Schwarzschild spacetime with respect to a particular component of ``infinity" (left), contrasted with the black region $\mathscr{B}$ (right).}
\label{fig:deSitterSchwarzschild2}
\end{figure}

\begin{figure}[b!]
\centering
\includegraphics[width=13cm]{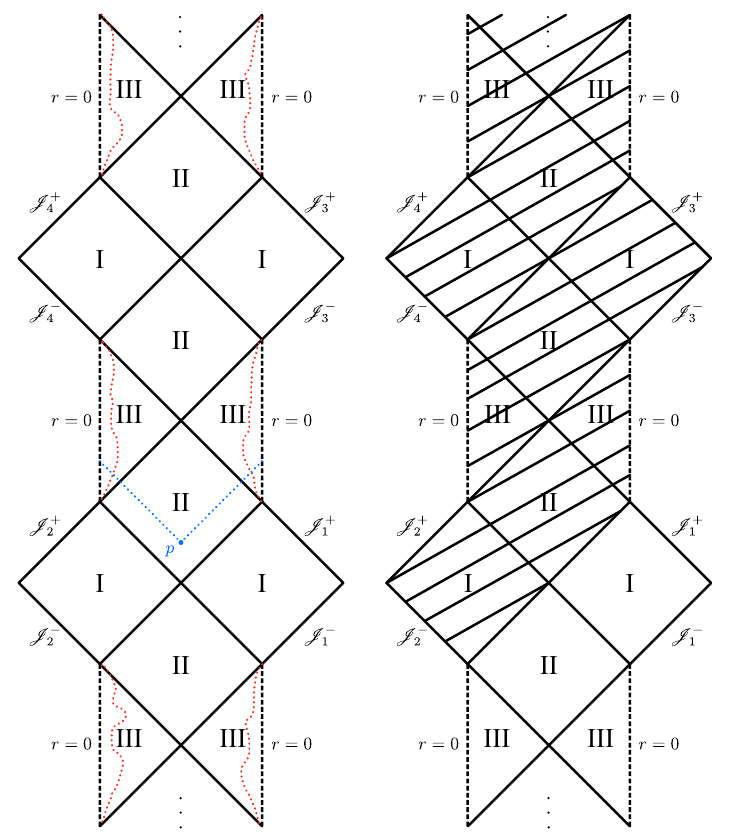}
\caption{\footnotesize Identifying the black region in the Kerr spacetime. The classic black region $\mathscr{B}_c$ with respect to $\mathscr{J}_1^+$ is shown on the right.}
\label{fig:kerr}
\end{figure}

We discuss the application of Definition \ref{definition:blackregion} to some standard black hole spacetimes, temporarily identifying singular neighborhoods in a heuristic fashion for conceptual clarity. In each of the following, ``singularities" are simply identified via inextendible, incomplete geodesics in maximal spacetimes, given a point-set and topological structure via standard coordinate charts in which these geodesics have endpoints. We present these examples prior to providing a more precise, coordinate-invariant formulation of singular neighborhoods in Section \ref{section:singularnbhd} to emphasize that there are multiple ways one might attempt to rigorously characterize singular neighborhoods, and that the singular neighborhoods depicted in Figures \ref{fig:singularnbhd} through \ref{fig:kerr} should be taken as the underlying ideal of what we would {\it like} the term to mean.

As a zeroth example, we comment that a trivial class of examples are the aformentioned ``nonsingular" spacetimes, which admit no singularly compact future sets, and hence have empty black region $\mathscr{B}$, under mild causality constraints (say, a distinguishing condition-- see Lemma \ref{lemma:distinguishingcausalcurve}).

\begin{example}\label{example:schwarzschild}
$M = $ the Schwarzschild spacetime.
\end{example}
Using Lemma \ref{lemma:blackcausalfuture}, we would like to identify those points in the Schwarzschild spacetime which are in the black region $\mathscr{B}$. We ask, then, which points $p \in M$ have the property that their causal future is singularly compact. This analysis is represented pictorially in Figure \ref{fig:schwarzschildEx}: we consider points $p_1$ in region I and $p_2$ in region II (the subset commonly identified as the black hole). The critical observation to make is that, upon removing from $J^+(p_2) = \overline{J^+(p_2)}$ any singular neighborhood (such as $U$, demarcated by the red dotted curves), it clearly becomes compact, being closed and bounded in the plane of the Penrose diagram. On the other hand, the singular neighborhood $U$ is an example of one for which $J^+(p_1) \backslash U$ is noncompact-- even upon the removal of $U$, there remain sequences in $J^+(p_1)$ which limit to the future null infinity $\mathscr{J}^+_1$, which is not in the spacetime.

These observations demonstrate that $p_2 \in \mathscr{B}$, while $p_1 \notin \mathscr{B}$. Clearly the same reasoning as for $p_2$ applies to any point in region II, while the same reasoning as for $p_1$ applies to any point in either asymptotically flat region (I and IV) or the white hole region (III). Hence region II is a subset of $\mathscr{B}$, while any point outside of region II is not in $\mathscr{B}$. The boundary of region II in $M$ is then identified as the event horizon $H = \partial \mathscr{B}$ as expected, and points in this boundary are not included in $\mathscr{B}$ according to how we have drawn $U$ (not necessarily enveloping the timelike infinities).

This picture aligns with expectations quite well, but it is worth distinguishing it from the standard paradigm. The sequence of definitions in Section \ref{section:standard} yields that there are two distinct classic black hole regions $\mathscr{B}_c$, according to whether one chooses $\mathscr{J}_1^+$ or $\mathscr{J}_2^+$ as the preferred notion of infinity. With respect to $\mathscr{J}_1^+$, the black hole region is comprised of regions II and IV; with respect to $\mathscr{J}_2^+$, it is comprised of regions II and I. In contrast, Definition \ref{definition:blackregion} identifies a single black region $\mathscr{B}$, comprised of region II alone-- the (interior of the) intersection of the two classic black hole regions. See Figure \ref{fig:schwarzschildEx2}. Of course, dual statements hold for white holes. \\
\qed

\begin{example}\label{example:deSitterSchwarzschild}
$M =$ the deSitter Schwarzschild spacetime.
\end{example}
We study the deSitter Schwarzschild spacetime, a minimalistic model of a black hole situated within an expanding universe, effectively Schwarzschild plus a positive cosmological constant. Shown in Figure \ref{fig:deSitterSchwarzschild} is this spacetime's maximally extended Penrose diagram, with points $p_1$ and $p_2$ in the regions II$_c$ and II$_b$, respectively. The subscript $c$ denotes those regions near ``cosmological" infinity, outside the cosmological horizons centered around the Schwarzschild-like regions II$_b$ and III$_w$. A first observation is that this spacetime is definitively not asymptotically flat due to the presence of the cosmological constant, which manifests in the Penrose diagram via the boundary apparently at ``infinity" being spacelike instead of null. While the conceptual picture laid out in Section \ref{section:standard} for identifying $\mathscr{B}_c$ may be applied in spirit since there seem to be natural candidates for pieces of ``future infinity" in this spacetime, the technical details may not, and standard results surrounding $\mathscr{B}_c$ therefore cannot be directly applied. Proceeding heuristically yields many choices of what one might mean by the black hole region $\mathscr{B}_c$, one for each component of $r= \infty$, and each instantiation of $\mathscr{B}_c$ contains every other cosmological infinity. See Figure \ref{fig:deSitterSchwarzschild2}.

The identification of $\mathscr{B}$, meanwhile, follows the Schwarzschild analysis rather directly. Indeed, looking at the causal future of $p_2$, it is still the case that it becomes compact upon removing the typical singular neighborhood shown in red, so $p_2 \in \mathscr{B}$. Removing the same from $J^+(p_1)$, there remain sequences in $J^+(p_1)$ which approach the $r=\infty$ boundary, so $p_1 \notin \mathscr{B}$. As before, these reasonings can easily be extended to find that the regions of type II$_b$ are all contained in $\mathscr{B}$, while the regions of type I, II$_c$, III$_c$, III$_w$, and IV are all in the complement of $\mathscr{B}$. Hence, $\mathscr{B}$ is precisely the union of all regions of type II$_b$, as one would expect.

Definition \ref{definition:blackregion}, then, is naturally able to accommodate black holes to which the standard paradigm does not directly apply and give the expected results, with no global structure required.

\qed

\begin{example}\label{example:kerr}
$M = $ the Kerr spacetime.
\end{example}
Having seen a successful deviation of Definition \ref{definition:blackregion} from the standard paradigm with the deSitter Schwarzschild spacetime, we now turn to one which may be seen as questionable in the maximally extended Kerr spacetime. See the usual (subextremal) Penrose diagram of a $\theta = \frac{\pi}{2}$ slice in Figure \ref{fig:kerr}. Depicted there is a point $p$ in a region of type II, with the boundary of its causal future shown as blue dotted lines. We notice that, though $p$ is in the classic black hole region $\mathscr{B}_c$ with respect to $\mathscr{J}_1^+$ (shown on the right) and $\mathscr{J}_2^+$, it is {\it not} with respect to $\mathscr{J}_3^+$ and $\mathscr{J}_4^+$. In fact, it is the case in maximally extended Kerr that every point in $M$ is in the past of the future null infinity associated to some asymptotically flat region, in contrast to region II in Schwarzschild (Figure \ref{fig:schwarzschildEx}) and regions II$_b$ in deSitter Schwarzschild (Figure \ref{fig:deSitterSchwarzschild}). Since $\mathscr{B}$ is heuristically supposed to be the intersection of all possible $\mathscr{B}_c$, this means that $\mathscr{B}$ should be empty!

Indeed, looking at the causal future of $p$ (as for any other point in $M$), we see that upon the removal of the typical singular neighborhood shown (red dotted curves), there still remains a sequence limiting to a sufficiently ``late" $\mathscr{J}^+$. This indicates $J^+(p)$ is not singularly compact, and so $p \notin \mathscr{B}$, and $\mathscr{B}$ is empty. That is, there is no black region in the Kerr spacetime according Definition \ref{definition:blackregion}. This should not be entirely surprising, as there is apparently no universal sense in which any classic black hole region $\mathscr{B}_c$ is small-- it always contains the entirety of infinitely many asymptotically flat regions.

Should one conclude, then, that Definition \ref{definition:blackregion} is incompatible with the notion of a rotating black hole? Given that the prevailing hope amongst physicists (encoded in Penrose's Strong Cosmic Censorship Conjecture \cite{christodoulou1999global, dafermos2014black, luk2019strong, penrose1979singularities}) is that the full structure of a maximal, dynamically forming black hole spacetime, rotating or no, is more similar to Schwarzschild's Penrose diagram than to Kerr's (in particular, it is expected to be globally hyperbolic), one would still expect such a spacetime to have a nonempty black region 
$\mathscr{B}$. That is, Definition \ref{definition:blackregion} should be able to characterize generic physical rotating black hole solutions, given strong cosmic censorship, even though it deems Kerr itself to have an empty black region. This is not too unreasonable a caveat given that the classic black hole region $\mathscr{B}_c$ of Section \ref{section:standard} is typically only even defined at all under the assumption of weak cosmic censorship. In any event, this is an inescapable peculiarity of taking the intuitive description ``light cannot escape" entirely seriously: there simply is no global sense in which this true for any portion of Kerr without invoking a preferred infinity, which we cannot easily do under a program that remains applicable outside of asymptotic flatness. It is up to the reader to decide whether this captures the physical phenomenon they wish to consider in a given scenario. We would argue at the very least that there  exist some considerations, including the framing of weak cosmic censorship, under which Definition \ref{definition:blackregion} captures precisely what one would want irrespective of the peculiarity encountered here.

\qed

\section{What is a Singular Neighborhood?} \label{section:singularnbhd}

\subsection{Avoiding Pathologies}
At the core of the proposed characterization of black holes is a notion of ``where" the singularities in a spacetime $(M,g)$ lie through the concept of a singular neighborhood. One means of obtaining this notion is heuristically through looking at the structure of $\partial M$ in particularly attractive coordinate charts or embeddings, such as Penrose diagrams, in which it seems intuitively clear where the manifest singularities are and adopting the embedding's or coordinates' topology. This is precisely the program carried out in the examples of Section \ref{section:blackhole}, and it is similar to the route one typically takes in trying to employ the intuition of the standard black hole paradigm where it isn't technically applicable (e.g.\@ discussing the past of infinity in deSitter-Schwarzschild). This may well give reasonable, desirable, and intuitive notions in many examples, and as such is not a universally incorrect approach. However, it is neither entirely generalizable nor invariant: the conclusions one makes may well depend upon the choice of coordinates or embedding, and it is a rather tricky prospect to characterize a ``correct" or ``optimal" choice, and even trickier to establish one's existence.

We take the perspective that singular neighborhoods should entirely be properties of $(M,g)$, with no extrinsic structure or constricting hypotheses required to make sense of them (and hence of black holes). The boundary constructions mentioned in the introduction seemingly provide a natural means of defining singular neighborhoods in accordance with this perspective: they require nothing more than $M$ being a smooth manifold and $g$ being a sufficiently smooth Lorentzian metric-- typically even time orientability isn't strictly required. Singularities are subtle, however: a critical topological shortcoming that arises in many such constructions, including the b- and g-boundaries, is that the singularities in $\partial M \subset \overline{M}$ are not guaranteed to be Hausdorff-separated from points in $M$. Indeed, Geroch, Can-Bin, and Wald \cite{geroch1982singular} demonstrated that any construction sufficiently similar to the g-boundary will have this pathology, and it seems this result was one of the primary reasons that the community's interest in boundary constructions waned after the 1970's. 

In the language of singular neighborhoods, this pathology means that an intuitive stipulation we put forward previously, that singular neighborhoods include all complements of compact sets, is not necessarily satisfied when using such constructions to identify singular neighborhoods. This is undesirable, as compact sets are in a sense bounded in the interior of $M$, and so should not be arbitrarily close to the boundary $\partial M$. While the construction of black holes put forward in the previous section could, in principle, be carried out in spite of this pathology, we consider it to violate the physical intuition underlying the definition. For this reason, we consider the later-developed a-boundary, or abstract boundary, to be the most natural means of identifying singular neighborhoods, as its topological structure manifestly separates interior and boundary points.

\subsection{Singular Neighborhoods Via the Abstract Boundary}

In this and the following section, we draw heavily upon abstract boundary concepts, terminology, and notation to provide and work with a definite meaning of singular neighborhoods. A streamlined review of all we will need, as well as references to various works on the abstract boundary, is provided in the appendix. Assuming familiarity with that material, we proceed to denote by $\mathcal{S}_p(M) \subset \mathcal{B}(M) \subset \overline{M}$ the collection of pure singularity abstract boundary points, and by $\mathcal{I}(M) \subset \mathcal{B}(M) \subset \overline{M}$ the collection of abstract boundary points at infinity.

\begin{definition}\label{definition:singularneighborhood}
An open set $U \subset M$ will be called a \ul{singular neighborhood} provided that every pure singularity abstract boundary point is strongly attached to $U$. Equivalently, provided that $U = \widetilde U \cap M$ for some neighborhood $\widetilde U \subset \overline{M}$ of $\mathcal{S}_p(M)$ in $\overline{M}$ endowed with $\mathcal{T}_{sap}(M)$.
\end{definition}

Given the results and discussion of the appendix, this captures precisely that $U$ ``surrounds" any pure singularities as much as possible from within $M$, and so agrees with the intuition and heuristic descriptions put forward in Section \ref{section:blackhole}. Indeed, the following corollary to Proposition \ref{proposition:envelopmentembedding} demonstrates how this definition yields the picture of singular neighborhoods presented in the figures of that section:

\begin{corollary}\label{corollary:envelopmentsingularneighborhood}
If $U\subset M$ is a singular neighborhood and $\phi: M \to \widehat M$ is an envelopment, then there exists an open set $\widehat U \subset \widehat M$ containing all pure singularity boundary points in $\widehat M$ with $\widehat U \cap \phi(M) = \phi(U)$.
\end{corollary}

\proof
By definition, we may find an open neighborhood $\widetilde U \subset \overline M$ of $\mathcal{S}_p(M)$ satisfying $\widetilde U \cap M = U$. By Proposition \ref{proposition:envelopmentembedding}, $V := \overline \pi^{-1}(\widetilde U)$ is open in $\overline{\phi(M)}$, so there is an open set $\widehat U \subset \widehat M$ satisfying $\widehat U \cap \overline{\phi(M)} = V$. By definition of $\overline \pi$ on $\phi(M)$, we find
$$\widehat U \cap \phi(M) = V \cap \phi(M) = \phi(\widetilde U \cap M) = \phi(U).$$
Furthermore, $V \subset \widehat U$ contains $\overline \pi^{-1}(\mathcal S_p(M))$, which is precisely the collection of pure singularity boundary points in $\widehat M$ by definition of $\overline \pi$ on $\partial \phi(M)$.

\qed

This definition, in furnishing us with a rigorous description of the collection $\mathscr{U}$ of singular neighborhoods, now puts us in a position to prove some expected features of the other objects defined in Section \ref{section:blackhole}, singularly compact sets and the black region. First, we firmly establish the ``closeness" of singular neighborhoods to singularities with a straightforward topological result. 

\begin{proposition}\label{proposition:singlimset}
A subset $S \subset M$ which meets every singular neighborhood has a pure singularity as an accumulation point in $\overline M$.
\end{proposition}

\proof
We prove the contrapositive. Take $S \subset M$, and suppose it has no pure singularity as an accumulation point in $\overline{M}$. Then for each $s \in \mathcal{S}_p(M) \subset \mathcal{B}(M)$, we may find an open set $\widetilde U_s \subset \overline{M}$ of $s$ such that $S \cap \widetilde U_s$ is at most finite. Since $s \in \mathcal{B}(M)$ is $T_2$-separated from points in $M$, we may in fact take $S \cap \widetilde U_s$ to be empty. Then 
$$\widetilde U := \bigcup_{s \in \mathcal{S}_p(M)} \widetilde U_s$$
is an open set in $\overline{M}$, from which $S$ is disjoint, containing every pure singularity. By definition, then, $\widetilde U \cap M$ is a singular neighborhood which $S$ does not meet.

\qed

This ensures, for example, that any non-compact subset $A \subset M$ which is singularly compact, with respect to the singular neighborhoods induced by the abstract boundary, must have a pure singularity as an accumulation point in $\overline{M}$, in alignment with expectations. Indeed, it ensures that {\it every} sequence in $A$ without accumulation points in $M$ must have a pure singularity as an accumulation point in $\overline{M}$. This makes precise claims made in Section \ref{section:blackhole} regarding ``nonsingular" spacetimes (in particular, that their singularly compact sets are precisely their compact sets), which we can now describe as spacetimes which admit no pure singularity abstract boundary points. When working with geodesics as the b.p.p.\@ family of curves $\mathcal C$ (Definition \ref{definition:bpp}), for example, this includes all geodesically complete spacetimes. We'll find further use for Proposition \ref{proposition:singlimset} in the subsequent section. Another result of great interest to the physical merit of the singular neighborhoods induced by the abstract boundary can currently only be put forward as a conjecture:

\begin{conjecture}\label{conjecture:separation}
Under physically relevant choices of curve families $\mathcal{C}$ on a smooth Lorentzian manifold $(M,g)$ (e.g.\@ geodesics with affine parameter, $C^1$ curves with generalized affine parameter, the causal subfamilies of either of these, etc.), points in $\mathcal{I}(M)$ and $\mathcal{S}_p(M)$ are $T_2$-separated from each other.
\end{conjecture}

This conjecture would ensure that abstract boundary points at infinity cannot be topologically embroiled with pure singularities (in particular, it is equivalent to the claim that there cannot exist a sequence in $M$ which limits to both a pure singularity and a point at infinity). It is easily seen to be true that such points are $T_1$-separated, as $T_1$-separation is equivalent to neither of the points covering the other, but $T_2$-separation, while a reasonable conjecture, is not so immediately obvious. A proof of this conjecture would go a long way toward establishing the physical reasonability of utilizing the abstract boundary to identify singular neighborhoods, as seen in its corollary, Conjecture \ref{conjecture:blackcomplement}, below.

\section{Characterizing the Black Region $\mathscr{B}$} \label{section:results}

We have now provided a precise notion of a singular neighborhood, and therefore made precise the black region of Definition \ref{definition:blackregion}. As seen in the appendix, however, the technical machinery required to do so was rather nontrivial, and so it appears a difficult task to navigate this machinery and characterize $\mathscr{B}$ in a given example, in principle requiring understanding the structure of every possible envelopment $\phi: M \to \widehat M$. In practice, one should be able to carry out a process very similar to the heuristic approach taken in the examples of Section $\ref{section:blackhole}$. In this section, we both establish and conjecture results to this effect, as well as results affirming the intuitive features expected of $\mathscr{B}$. We begin by recalling some useful, well-known lemmas.

\begin{lemma}
The following conditions on a curve $\sigma:[0,b) \to M$, with $0 < b \leq \infty$, are equivalent:
\begin{enumerate}[(i)]
\item For any compact set $K \subset M$, $\sigma$ eventually leaves and never returns to $K$.
\item For any sequence $(t_n)_{n=1}^\infty \subset [0,b)$ satisfying $t_n \to b$, $(\sigma(t_n))$ does not converge in $M$. 
\end{enumerate}
\end{lemma}

\proof For $(i) \implies (ii)$, suppose $\sigma(t_n) \to p \in M$. Then given any compact neighborhood $K \subset M$ of $p$, $\sigma(t_n)$ is eventually always in $K$, in contradiction to $(i)$.

For $(ii) \implies (i)$, note that if $(i)$ does not hold, then there must be a compact $K \subset M$ and a sequence $(t_n)_{n=1}^\infty \subset [0,b)$ with  $t_n \to b$ such that $\sigma(t_n) \in K$, so that $\left( \sigma(t_n) \right)$ must have a convergent subsequence, meaning $(ii)$ does not hold.

\qed

Recall that the coverse to condition $(i)$ above is commonly referred to in the literature by saying $\sigma$ is {\it partially imprisoned} in some compact $K \subset M$. A more restrictive condition is that $\sigma$ be {\it totally imprisoned} in some compact $K \subset M$, meaning $\sigma$ eventually enters and remains inside $K$. Any curve satisfying the above conditions, then, cannot be even partially imprisoned. When $M$ is strongly causal, the following ensures that we may utilize these properties for any inextendible causal curve:

\begin{lemma}\label{lemma:stronglycausalcurve}
Let $\sigma:[0,b) \to M$ be an inextendible causal curve. If strong causality is not violated on the closure of the image of $\sigma$, then $\sigma$ has property (ii) (and hence (i)) in the preceding lemma. That is, $\sigma$ cannot be partially imprisoned.
\end{lemma}

\proof Suppose there exists a sequence $(t_n)_{n=1}^\infty \subset [0,b)$ such that $t_n \to b$ and $\sigma(t_n) \to p \in M$. Since $\sigma$ is inextendible, there must exist a sequence $(s_n)$ satisfying $s_n \to b$ such that $(\sigma(s_n))$ does not converge to $p$, so there is a neighborhood $U$ of $p$ such that $\sigma(s_n)$ has a subsequence not meeting $U$. By passing to appropriate monotonic subsequences, we may assume $t_{n} \leq s_n \leq t_{n+1}$ and $(\sigma(s_n))$ does not meet $U$. For any neighborhood $V \subset U$ of $p$, however, $(\sigma(t_n))$ eventually enters and remains inside $V$, so that the causal curves $\sigma|_{[t_n,t_{n+1}]}$ violate strong causality at $p$. This is a contradiction since $p$ is in the closure of the image of $\sigma$.

\qed

A similar result can be obtained by weakening both the hypothesis and conclusion. We refer the reader to \cite{hellis} for proof:

\begin{lemma}[\cite{hellis}, Proposition 6.4.8]\label{lemma:distinguishingcausalcurve}
Let $\sigma:[0,b) \to M$ be an inextendible causal curve. If either the past or future distinguishing conditions holds on a compact set $K \subset M$, then $\sigma$ cannot be totally imprisoned in $K$.
\end{lemma}

\qed

In particular, these results affirm that standard compactness has no hope of describing causally well-behaved black holes, as they indicate that any nonempty future set cannot be compact in a causally distinguishing spacetime (since any such set contains inextendible causal curves). With them, however, we may establish the singular nature of $\mathscr{B}$ under such causality conditions.

\begin{proposition}\label{proposition:singlimcurve1}
If an inextendible, future-directed causal curve $\sigma:[0,b) \to M$ which meets the black region $\mathscr{B}$ is not totally imprisoned, then $\sigma$ has a pure singularity accumulation point in $\overline{M}$.
\end{proposition}

\proof 
Let $\sigma:[0,b) \to M$ be an inextendible, future-directed causal curve meeting $\mathscr{B}$, i.e. meeting some singularly compact set $A \subset M$  satisfying $J^+(A) = A$ (so $\sigma$ remains in $A$), and suppose $\sigma$ is not totally imprisoned. For each singular neighborhood $U \in \mathscr{U}$, $A \backslash U$ is compact. Since $\sigma$ is not totally imprisoned, then, there exists a sequence $(t_n) \subset [0,b)$ with $t_n \to b$ such that $\sigma(t_n)$ is not in $A \backslash U$, so we must have $\sigma(t_n) \in U$. In particular, $\sigma$ meets every singular neighborhood. Taking $S$ to be the image of $\sigma$ in Proposition \ref{proposition:singlimset} completes the proof.

\qed

Strengthening the hypothesis by switching total imprisonment to partial imprisonment, we can get a more compelling result:

\begin{proposition}\label{proposition:singlimcurve2}
If an inextendible, future-directed causal curve $\sigma:[0,b) \to M$ which meets the black region $\mathscr{B}$ is not partially imprisoned, then every sequence along the end of $\sigma$ has a pure singularity accumulation point in $\overline{M}$.
\end{proposition}

\proof 
Let $\sigma:[0,b) \to M$ be an inextendible, future-directed causal curve meeting $\mathscr{B}$, i.e. meeting some singularly compact set $A \subset M$  satisfying $J^+(A) = A$ (so $\sigma$ remains in $A$), and suppose $\sigma$ is not partially imprisoned. For each singular neighborhood $U \in \mathscr{U}$, $A \backslash U$ is compact. Since $\sigma$ is not partially imprisoned, then, there is a $t_U \in [0,b) $ such that $\sigma(t) \notin A \backslash U$, and hence $\sigma(t) \in U$, for $t > t_U$. Thus, given any sequence $(t_n)_{n=1}^\infty \subset [0,b)$ with $t_n \to b$, $(\sigma(t_n))$ meets every singular neighborhood $U  \in \mathscr{U}$, and the result follows from Proposition \ref{proposition:singlimset}.

\qed

Combining either of these with the previous lemmas, then, yields the desired singular nature of $\mathscr{B}$. We only state the strongly causal conclusion in the following, but it should be clear that the same statement is true under the distinguishing condition upon the removal of the phrase ``sequence along the end of an" from the theorem.

\begin{theorem}\label{theorem:singlim}
Let $(M,g)$ be strongly causal. If $p \in \mathscr{B}$, then every sequence along the end of an inextendible, future-directed causal curve through $p$ has a pure singularity as an accumulation point in $\overline{M}$.

 \qed
\end{theorem}

\begin{figure}[t!]
\centering
\includegraphics[width=10cm]{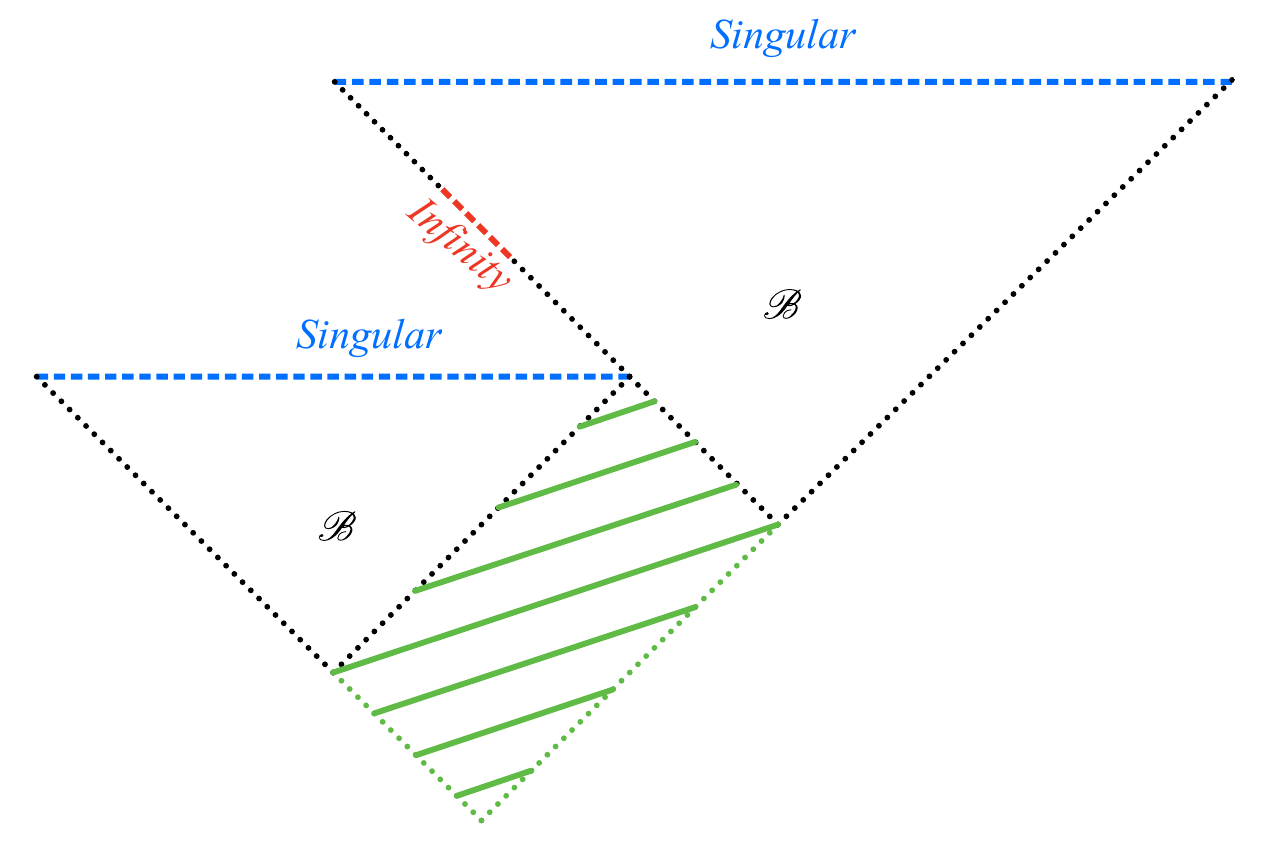}
\caption{ \footnotesize A schematic counterexample to the converse of Theorem \ref{theorem:singlim}. The dashed lines are excised from the spacetime, which has the causal structure of $2$D Minkowski space. Under the envelopment of the figure, the excised points are taken to be pure singularities (blue dashed lines) and points at infinity (red dashed line). The black dotted lines are null curves, some of which comprise the boundary of $\mathscr{B}$. The green shaded region is the set of points which satisfy the consequent of Theorem \ref{theorem:singlim} but not the antecedent. These points are not in $\mathscr{B}$ due to the presence of the points at infinity, but every inextendible, future-directed causal curve through them must approach a pure singularity. This example is not explicit, though we believe one can construct such a spacetime by excising the dashed lines from Minkowski space and introducing a conformal factor which appropriately diverges to $0$ or $\infty$ on them.}
\label{fig:counterexample}
\end{figure}

It is in this sense that the black region $\mathscr{B}$ satisfies the heuristic description of being in the ``past Cauchy development of the singularities": any future-directed causal curve emanating from a point in the black region must approach a pure singularity in the abstract boundary, under reasonable causality restrictions. It should be expected, of course, that such conditions are needed, as causality conditions are what put structure on objects like $J^+(p)$ for some $p \in M$, and $\mathscr{B}$ is built using such causal objects. The converse to this theorem is not strictly true-- see the example in Figure \ref{fig:counterexample}. 

A partial converse to the heuristic, that any point in the ``past Cauchy development of the singularities" should indeed lie in the black region, can be obtained, however. We provide this in the following two theorems. The first is a conceptual result to this effect at the level of the full structure of the abstract boundary, which both supplies this heuristic and links the present notion of a black hole to the standard perspective.

\begin{theorem}\label{theorem:globallyhyperbolicB}
Suppose $(M,g)$ is globally hyperbolic, take the b.p.p.\@ family of curves $\mathcal{C}$ to be $C^1$ curves with generalized affine parameter, and take $p \in M$. If there are no points at infinity strongly attached to $I^+(p)$, then $I^+(p) \subset \mathscr{B}$ (and hence $p \in \overline{\mathscr B}$).
\end{theorem}

\proof
Take $\Sigma \subset M$ to be a Cauchy hypersurface for $M$. It is a classic result originally due to Geroch \cite{gerochgloballyhyperbolic} that there exists a homeomorphism, strengthened to a diffeomorphism by Bernal and Sanchez \cite{bernalsanchez}, $\phi: M \to \mathbb{R} \times \Sigma$ satisfying
\begin{enumerate}[(i)]
\item $\Sigma_t := \phi^{-1}\left(\{t\} \times \Sigma \right)$ is a Cauchy hypersurface in $M$ for each $t \in \mathbb{R}$.
\item For each $x \in \Sigma$, $t > s \implies \phi^{-1}(t,x) \in I^+ \left( \phi^{-1}(s,x) \right)$.
\item The ``time function" $\tau : M \to \mathbb{R}$, defined as $\tau := \pi_1 \circ \phi$ with $\pi_1: \mathbb{R} \times \Sigma \to \mathbb{R}$ the projection onto the first coordinate, is strictly increasing on future-directed causal curves.
\end{enumerate} Denoting for $t \in \mathbb{R}$
$$ \Sigma_t^- := D^-(\Sigma_t) = \bigcup_{s \leq t} \Sigma_s,$$
it is also a classic result that $J^+(p) \cap \Sigma_t^-$ is compact (\cite{oneil}, Lemma 14.40). 

Similarly to $\tau$, define the smooth map $\rho: M \to \Sigma$ by $\rho := \pi_2 \circ \phi$, where $\pi_2$ is the projection onto the second coordinate, and set 
$$C_t := \rho \left( I^+(p) \cap \Sigma_t \right) \subset \Sigma.$$
By property (ii), $s < t \implies C_s \subset C_t$, which together with property (iii) implies that $C_t$ is path-connected. Indeed, if $q \in I^+(p)$, there exists a timelike curve $\sigma:[0,1] \to M$ from $\sigma(0) = p$ to $ \sigma(1)=q$, and $\rho \circ \sigma$ is a continuous path from $\rho(p)$ to $\rho(q)$ in $C_{\tau(q)}$, since $\rho \circ \sigma(s) \in C_{\tau(\sigma(s))} \subset C_{\tau(q)}$ for each $s \in (0,1]$. This shows every point in $C_t$, for $t > \tau(p)$, can be connected via a path in $C_t$ to $\rho(p)$. For $t \leq \tau(p)$, $C_t$ is empty.

Consider a sequence $(p_n)_{n=1}^\infty \subset I^+(p)$ with no accumulation points in $M$. Since $J^+(p) \cap \Sigma_t^-$ is compact, for each $t \in \mathbb{R}$ we must eventually have $\tau(p_n) > t$. Hence we may assume, perhaps upon passing to a subsequence, that $t_n := \tau(p_n)$ is strictly increasing with $t_n \to \infty$. We can now construct a curve $\gamma$ from $p_1$ to $p_2$ by using condition (ii) to follow a future-directed timelike curve from $p_1 \in \Sigma_{t_1}$ to $\phi^{-1}(t_2,\rho(p_1)) \in \Sigma_{t_2}$, and then use the path-connectedness of $C_{t_2}$ to travel within $I^+(p) \cap \Sigma_{t_2}$ to $p_2$. Iterating this procedure, we extend $\gamma$ to a curve through the sequence $(p_n)$, entirely contained in $I^+(p)$, on which $\tau \to \infty$ and is non-decreasing. The last condition ensures that $\gamma$ has no accumulation points in $M$, as any sequence along the end of $\gamma$ has $\tau \to \infty$. Smoothing out $\gamma$ under these conditions, the endpoint theorem for curves \cite{scott2021endpoint}\footnotemark{} ensures $\gamma$, and hence $(p_n)$, limits to an approachable abstract boundary point $[p_\infty] \in \mathcal{B}(M)$ which is strongly attached to $I^+(p)$. By hypothesis, $[p_\infty]$ can neither be nor cover a point at infinity, and hence it must be a pure singularity. $[p_\infty]$ is then strongly attached to every singular neighborhood by definition, so $(p_n)$ must meet every singular neighborhood.

Now, take $p_0 \in I^+(p)$, and fix a singular neighborhood $U \subset M$. The above argument demonstrates that any sequence $(p_n)_{n=1}^\infty \subset I^+(p) \backslash U$ must have an accumulation point in $M$, and hence that $\overline{I^+(p) \backslash U}$ is compact. Since $M$ is globally hyperbolic, it is causally simple, so that $J^+(p_0)$ is closed. Hence $J^+(p_0) \backslash U \subset I^+(p) \backslash U$ is a closed subset of $\overline{I^+(p) \backslash U}$, so it is compact. This demonstrates $J^+(p_0) = \overline{J^+(p_0)}$ is singularly compact, and therefore that $p_0 \in \mathscr{B}$ by Lemma \ref{lemma:blackcausalfuture}.

\qed

\footnotetext{The {\it endpoint theorem for curves} refers to the result that any smooth, non-self-intersecting curve in $M$ without accumulation points limits to some abstract boundary point in $\overline{M}$. It is a trivial consequence of the proof that if the curve can be constrained to lie entirely in some open set, the provided endpoint can be chosen to be strongly attached to that open set.}

Noting that $I^+(p)$ having no points at infinity strongly attached is equivalent to saying that every abstract boundary point strongly attached to $I^+(p)$ is a pure singularity, as well as that the endpoint theorem for curves ensures that an inextendible causal curve contained in an open set $V \subset M$, with $M$ strongly causal, limits to an abstract boundary point in $\overline{M}$ strongly attached to $V$, the hypotheses of this theorem ensure that any inextendible causal curve starting at $p_0 \in I^+(p)$ must limit to a pure singularity, i.e.\@ that $p_0$ is in the ``past Cauchy development of the singularities". Under this hypothesis, the theorem establishes that $p_0 \in \mathscr{B}$, as claimed in the heuristic. 

Moreover, Theorem \ref{theorem:globallyhyperbolicB} indicates that $\mathscr{B}$, when formalized via the abstract boundary, is closely related to the complement of the past of infinity, the more standard means of defining a black hole. This is a testament both to the reasonable nature of $\mathscr{B}$ and to the naturalness of the structure of the abstract boundary. Recall that the original definition of $\mathscr{B}$, Definition \ref{definition:blackregion}, makes no mention of any concept of infinity, nor even requires that one exists-- the abstract boundary formalism furnished both the concept of infinity in a general spacetime and tied that concept to the black region. A priori, there was no reason that such a link must hold.

The weakness of Theorem \ref{theorem:globallyhyperbolicB} is twofold, however. The first weakness is the somewhat restrictive causality condition it requires. We expect a similar result is true in, say, a causally simple setting, but a proof remains elusive. The second is in the difficulty of applying Theorem \ref{theorem:globallyhyperbolicB} in practice, as making a conclusion about all possible abstract boundary points strongly attached to an open set seems rather difficult to do. Fortunately, the next result (though it says much less about how $\mathscr B$ is related to the full abstract boundary structure) ameliorates both of these concerns by providing a more practical means of identifying points in $\mathscr{B}$, free of causality constraints or a restriction on $\mathcal{C}$.

\begin{theorem}\label{theorem:compactblack}
If there exists an envelopment $\phi: M \to \widehat M$ under which $\overline{\phi(I^+(p))}=\overline{\phi \left( \overline{J^+(p)} \right)}$ is compact and every boundary point in $\widehat{M}$ attached to $I^+(p)$ is a pure singularity, then $p \in \mathscr{B}$.
\end{theorem}

We provide two proofs of distinctly different flavors, one sequential and one more directly wielding the topological machinery of the abstract boundary.
\\

{\it Proof 1.} Let $U \subset M$ be a singular neighborhood, and consider a sequence $(p_n)_{n=1}^\infty \subset \overline{J^+(p)} \backslash U.$ Since $\overline{\phi(I^+(p))}$ is compact, $\left( \phi(p_n) \right)$ has an accumulation point $q \in \overline{\phi(I^+(p))} \subset \overline{\phi(M)}$. Suppose that $q \in \partial\phi(M)$. Being attached to $I^+(p)$, $q$ would then be a pure singularity by hypothesis, and by Definition \ref{definition:singularneighborhood}, $q$ would be strongly attached to $U$. This is a contradiction, however, since $(p_n)$ does not meet $U$. Thus we must have that $q \in \phi(M)$, and so $\phi^{-1}(q)$ is an accumulation point of $(p_n)$ in $M$. Since $\overline{J^+(p)} \backslash U$ is closed, this shows $\overline{J^+(p)} \backslash U$ is compact.

This demonstrates that $\overline{J^+(p)}$ is singularly compact, and therefore that $p \in \mathscr{B}$ by Lemma \ref{lemma:blackcausalfuture}.

\qed

{\it Proof 2.} Let $U \subset M$ be a singular neighborhood. By Corollary \ref{corollary:envelopmentsingularneighborhood}, there exists an open set $\widehat U \subset \widehat M$ containing all pure singularities in $\widehat M$ and satisfying $\widehat U \cap \phi(M) = \phi(U)$.
Since every boundary point in $\widehat M$ attached to $I^+(p)$ is a pure singularity, we have
 $$\overline{\phi ( I^+(p) )} \backslash \phi \left( \overline{J^+(p)} \right) = \overline{\phi(I^+(p))} \cap \partial \phi(M) \subset \widehat U.$$
Hence we find 
$$\overline{\phi(I^+(p))} \backslash \widehat U = \phi \left( \overline{J^+(p)} \right) \backslash \phi(U) = \phi \left( \overline{J^+(p)} \backslash U \right).$$
Since $\widehat U$ is open and $\overline{\phi(I^+(p))}$ is compact by hypothesis, $\overline{\phi(I^+(p))} \backslash \widehat U$ is compact, and thus so is $\overline{J^+(p)} \backslash U$ by the above equality. This demonstrates $\overline{J^+(p)}$ is singularly compact.

\qed

This theorem provides the formal justification, when defining things precisely via the abstract boundary formalism, for taking the heuristic approach presented in the examples of Section \ref{section:blackhole} to identifying points in $\mathscr{B}$. It is by far the most useful result to identifying points in $\mathscr{B}$ in practice, in that it reduces the problem of investigating all possible abstract boundary points (a seemingly intractable task) to investigating only those boundary points arising in a particular envelopment.

The other side of identifying the subset $\mathscr{B}$, namely of identifying those points in $M$ which are {\it not} in $\mathscr{B}$, is more subtle. An approach is provided, in principle, by Theorem \ref{theorem:singlim}: if one could find a sequence along an inextendible, future-directed causal curve through a point $p \in M$ which does {\it not} have a pure singularity as a limit point, then $p \notin \mathscr{B}$. It is difficult to establish in practice, however, that a sequence in $M$ does not have a pure singularity as a limit point under ${\it any}$ envelopment without the topological constraint of Conjecture \ref{conjecture:separation}. To formalize this, we state the following conjecture which follows readily from Conjecture \ref{conjecture:separation}, and which would more firmly tie the present notion of a black hole to the standard perspective.

\begin{conjecture}\label{conjecture:blackcomplement}
If $p \in M$ has a point at infinity attached to $I^+(p)$, then $p \notin \mathscr{B}$.
\end{conjecture}

{\it Proof (provided Conjecture \ref{conjecture:separation}).} 
Suppose $p \in \mathscr{B}$ has a point at infinity $[q] \in \overline{M}$ attached to $I^+(p)$. Then by definition there is a sequence of points $(p_n) \subset I^+(p) \subset \overline{J^+(p)}$ such that $p_n \to [q]$ in $\overline{M}$. Since $[q]$ is $T_2$-separated from points in $M$, $(p_n)$ has no accumulation points in $M$, so it enters and remains inside of every singular neighborhood by the singular compactness of $\overline{J^+(p)}$. By Proposition \ref{proposition:singlimset}, $(p_n)$ has a pure singularity $[s] \in \overline{M}$ as an accumulation point, and hence passing to an appropriate subsequence yields both $p_n \to [q]$ and $p_n \to [s]$, in contradiction to Conjecture \ref{conjecture:separation}.

\qed

\section{Weak Cosmic Censorship}\label{section:weakcosmiccensorship}
Having formally defined a general notion of black hole applicable to any maximal spacetime, we turn to an important outstanding question for theoretical general relativity which centers around features of black holes, the Weak Cosmic Censorship Conjecture. The incompleteness theorems of Hawking and Penrose (\cite{hawking1965occurrence}, \cite{penrosesingularity}), early investigations of dynamical collapse by Oppenheider and Schneider \cite{oppenheimer1939continued}, and Schoen and Yau's demonstration of the dynamical nature of trapped surfaces \cite{schoen1983existence} have convinced physicists that singularities are a generic feature of general relativity which must be wrangled with. This is not so surprising, as it was known and expected that general relativity must give way to quantum corrections in extreme situations. On its face, however, it may spell disaster for the predictive power of the theory in that singularities, by definition, cannot be evolved through. Points to the future of a singularity would necessarily be causally influenced in a manner that could not be modeled within the context of general relativity, and hence the structure and dynamics of these points could not be predicted by it. The generic emergence of singularities essentially means, then, that classical physics is not even nominally self-contained. 

This situation might not be so disastrous, however, if the futures of singularities could be said to be ``small" in some sense. That is, this is not so big of a problem to the conceptual value of the theory so long as one can guarantee solvability sufficiently far from singularities. The Weak Cosmic Censorship Conjecture, due to Penrose \cite{penrose1969gravitational}, is the hope that this is the case, with its heuristic content being that {\it singularities in general relativity are generically hidden behind black holes}. This would mean that, at least generically, the futures of singularities must be contained in the ``small" interiors of black holes, and hence they do not pose a significant obstruction to general relativity's self-consistently modeling the universe at large. Singularities in violation are dubbed {\it naked} \footnote{Some authors distinguish between {\it locally} and {\it globally} naked singularities, respectively being violations of strong and weak cosmic censorship. We have described and exclusively refer to globally naked singularities.}.

Historically, this conjecture has been formulated as an initial value problem, a statement about spacetimes which evolve from initial data stipulated on a sufficiently nice (e.g.\@ complete, asymptotically flat) Riemannian 3-manifold. This was classically formalized \cite{hellis, waldgr} by positing that such a spacetime, the maximal Cauchy development of the initial data in question, generically turns out to be future asymptotically predictable (recall Definition \ref{definition:asymptoticallypredictable}). More modern formulations \cite{christodoulou1999global, rodnianski2019naked} are framed in terms of the ``completeness of future null infinity", which, while still requiring the notion of asymptotic flatness, avoids dependence on an explicit conformal embedding. The seminal results in this domain are Christodoulou's demonstration for a real scalar field matter source in spherical symmetry \cite{christodoulou1999instability} and Christodoulou and Kleinerman's demonstration for vacuum perturbations of Minkowski initial data \cite{christodoulou1993global}. Review and discussion of this topic can be found in \cite{christodoulou1999global, joshi2000gravitational, singh1996gravitational, wald1999gravitational}.

A significant challenge to achieving a comprehensive result to this effect is the differing characteristics of  various matter fields when coupled to the Einstein equation, hence results are generally restricted either to vacuum or particular matter fields. Moreover, it is not entirely clear what the ``correct" complete set of physical matter fields is, so the goalpost for a physically compelling resolution is difficult to set. Moreover, the typical formalization requires asymptotic flatness, though the physical significance of the conjecture still persists outside of this context. While it was originally conceived with isolated gravitational collapse in mind, the spirit of weak cosmic censorship is not only a physical problem for isolated systems -- this is manifest in the non-asymptotically flat cosmological structure of the universe at large, as well as in the consideration of singularities that may have arisen very early in the universe's history, in the cosmic primordial soup. 

We would like, then, to put forward a global formulation of this conjecture which depends critically neither on the particular choice of matter fields (perhaps only, say, energy condition restrictions on curvature) nor the assumption of asymptotic flatness to make sense. The following is a somewhat broad conjecture in this direction:

\begin{conjecture}\label{conjecture:globalcensor}
(Global Weak Cosmic Censorship) In a generic, maximal, physically admissible spacetime $(M,g)$ which admits a complete space-like hypersurface $\Sigma$, there exists a singular neighborhood $U \in \mathscr{U}$ such that $U \cap D^+(\Sigma) \subset \mathscr{B}$.
\end{conjecture}

Setting aside the hypotheses, this simply states that if a spacetime contains some ``full" instant of time $\Sigma$ at which there are no singularities, then no naked singularities will develop in that instant's future domain of dependence $D^+(\Sigma)$. Indeed, the condition's negation is that $D^+(\Sigma) \backslash \mathscr{B}$ meets  every singular neighborhood, which should mean (by Proposition \ref{proposition:singlimset} when formalizing via the abstract boundary) that $D^+(\Sigma) \backslash \mathscr{B}$ is arbitrarily close to a singularity-- that is, a singularity outside of the black region $\mathscr{B}$ develops within $D^+(\Sigma)$, precisely what weak cosmic censorship should avoid. This formulation in terms of some complete $\Sigma$ is necessary to rule out spacetimes which ``always" have a singularity in some sense, such as negative mass Schwarzschild, while still allowing for the initial singularity in cosmological models. We've dubbed our description ``Global" due to its working in terms of the full structure of a spacetime, rather than an initial data set to be evolved.

A helpful exercise to grok our terminology is observing that Conjecture \ref{conjecture:globalcensor} does {\it not} claim that singularities are generic. Recall from Section \ref{section:blackhole} that in a nonsingular spacetime, the collection of singular neighborhoods $\mathscr{U}$ is the entire topology (this agrees with the approach of Definition \ref{definition:singularneighborhood}: if there are no pure singularity abstract boundary points, then all of them are strongly attached to every open set in $M$), and in particular includes the empty set. Explicitly, then, Conjecture \ref{conjecture:globalcensor} is necessarily satisfied in a nonsingular spacetime by taking $U$ to be the empty set.

We now discuss the hypotheses. At this level of generality, the conjecture has several avenues along which it is subject to variation, as is familiar in dealings of cosmic censorship. The first is quite standard and common among all formulations: the relevant notion of ``generic" is rather flexible so long as it is reasonable. One might define a topology on the collection of appropriate spacetimes or on the space of appropriate metrics on a particular smooth manifold, and show the set of spacetimes or metrics satisfying the stated condition is open and its complement has empty interior. Alternatively, one might define a measure on the same spaces, and show that the set satisfying the condition has full measure. Some such genericness clause is necessary, as it is not difficult to write down maximal spacetimes which do not satisfy the conjectured condition while appearing physically admissible (see Example \ref{example:vaidya} and Figure \ref{fig:vaidya2} below).

The term ``maximal" is similarly subject to interpretation, depending upon the metric regularity class of interest. The metric's being $C^2$, time-orientable, and satisfying a causality condition are common physical invocations, though PDE approaches to IVP formulations lend themselves to weakening the $C^2$ condition to only requiring that the Christoffel symbols be in $L^2_\text{loc}$. Note that maximality is required for the claim to be meaningful, as it ensures that singular neighborhoods contain all relevant data-- clearly they cannot identify singular behavior which has been omitted from the spacetime by fiat (e.g., considering $M$ to be the region $r > m$ in Schwarzschild). One might assume something akin to strong cosmic censorship as a hypothesis to trivialize this condition.

``Physically admissible", also a standard fudge factor in cosmic censorship, should be taken as a substitute for the ad hoc restriction that one's matter fields admit a nice PDE description, with the hope being that, say, the dominant energy condition is all that is needed. It may well be that the conjecture cannot be proven true without taking this phrase to mean that the spacetime is a solution of Einstein's equation with a particular set of matter fields, but since it is unclear what the full suite of physical matter fields should be, it is desirable to have a formulation of the conjecture which at least allows for the possibility that this is not required.

Finally, as we have discussed at length, one might take several different approaches to rigorously defining singular neighborhoods, so this is another choice one can make in rendering Conjecture \ref{conjecture:globalcensor} a specific claim subject to proof. We have focused on one such framework through the abstract boundary, though there may well exist reasonable alternatives. Even within the context of the abstract boundary, one has some freedom in choosing the physically relevant b.p.p.\@ family of curves one uses to identify singularities. While $C^1$ curves of generalized affine parameter are the most comprehensive and compelling in view of Geroch's work \cite{geroch1968singularity}\footnote{Geroch constructed here a geodesically complete spacetime which contains an inextendible timelike curve of finite proper time and bounded acceleration, demonstrating that geodesic completeness is not entirely sufficient to rule out physical pathologies associated to singularities.}, geodesics are a natural starting point.

We conclude this section with an example of a family of simple spherically symmetric spacetimes containing naked singularities in an apparently generic fashion, at least within the considered family, to exhibit both the physical need for a formulation of weak cosmic censorship in the vein of Conjecture \ref{conjecture:globalcensor} as well as the challenge one must overcome in hoping to prove such a result.

\begin{example}\label{example:vaidya}
Naked singularities in Vaidya spacetimes.
\end{example}
It is well known that perhaps the simplest possible models for the dynamical formation of a black hole, the Vaidya spacetimes, can also demonstrate the formation of naked singularities \cite{joshi1992strong, kuroda1984naked, lemos1992naked}. We recall here our recent result \cite{wheelervaidya} indicating that naked singularities are significantly more generic within these models than was previously stressed in the literature. We consider the spacetime  $M= (\mathbb{R} \times \mathbb{R}^3) \backslash \{(v,0,0,0) \, | \, v \geq 0 \}$ with metric given by 
$$g=-\left(1-\frac{2m(v)}{r} \right) dv^2 + dv \otimes dr + dr \otimes dv + r^2 d\sigma^2, $$
where $d\sigma^2$ is the metric on the unit sphere $S^2 \subset \mathbb{R}^3$, $r$ is the radial coordinate on $\mathbb{R}^3$, and $m: \mathbb{R} \to \mathbb{R}_{\geq 0}$ is $C^2$, nondecreasing, and satisfies $m(v) = 0$ for $v \leq 0$, $m(v) > 0$ for $v > 0$, and $m(v) = m_0$ for $v \geq v_0> 0$. That is, the mass parameter increases $C^2$-smoothly from $0$ beginning at $v=0$, eventually leveling off to $m_0$ at $v=v_0$.

\begin{figure}[b!]
\centering
\includegraphics[width=16cm]{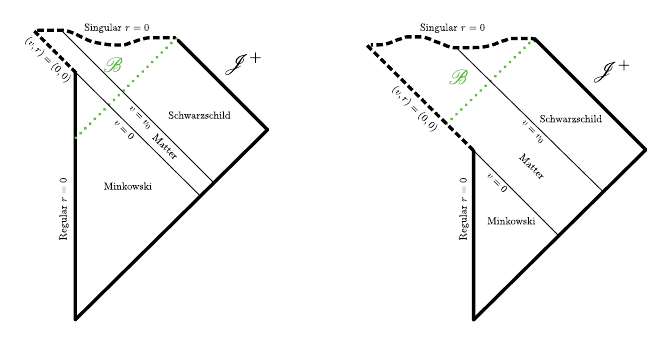}
\caption{ \footnotesize Penrose diagrams for two Vaidya spacetimes of interest, with smooth mass functions which grow at different rates. The Minkowski, Schwarzschild, and matter-containing regions are labelled in each case, separated by the lines of $v=0$ and $v=v_0$. The green dotted lines mark the boundaries of the black regions. When the mass function increases more slowly (on the right), the $(v,r)=(0,0)$ singularity, which is stretched into a diagonal line in null-null coordinates, becomes naked. See Proposition \ref{proposition:nakedvaidya}.}
\label{fig:vaidya1}
\end{figure}

The Einstein tensor has a single nonzero component in these coordinates-- it can be computed to be
\begin{equation}\label{equation:curvature}
G = \frac{2m'(v)}{r^2} dv^2,
\end{equation}
from which we easily deduce that the dominant energy condition is satisfied since $m'(v) \geq 0$. For $v \leq 0$, $g$ is precisely the Minkowski metric in ingoing null-radial coordinates, while for $v \geq v_0$, $g$ is precisely the Schwarzschild metric of mass $m_0$ in ingoing Eddington-Finkelstein coordinates, so this spacetime is a simple model of the formation of a Schwarzschild black hole of mass $m_0$ due to matter falling into the origin along the null geodesics of constant $0 < v < v_0$. Figure \ref{fig:vaidya1} gives a schematic Penrose diagram of two contrasting cases.

A case which has received much attention is the self-similar case, that of taking $m(v) = \frac{m_0}{v_0} v$ for $v \in [0,v_0]$ (this choice relaxes the regularity constraint to only holding on $[0,v_0]$), and it is well understood that in this scenario a naked singularity develops at $(v,r) = (0,0)$ provided that $\frac{m_0}{v_0} \leq \frac{1}{16}$. In fact, this occurs much more generally:

\begin{proposition}[\cite{wheelervaidya}, Proposition 3]\label{proposition:nakedvaidya}
Consider the Vaidya spacetime as above characterized by a nondecreasing mass function $m(v)$ which is $C^1$ on $[0,v_0]$ and satisfies $\sup_{0 < v < v_0} m'(v) \leq \frac{1}{16}$. There exists a one-parameter family (modulo $S^2$) of null geodesics which both terminate to the past at $(v,r) \to (0,0)$ and escape to $\mathscr{J}^+$ in the Schwarzschild region $v > v_0$.
\end{proposition}

\qed

The heuristic takeaway from this proposition is that if the in-falling matter accumulates at the origin {\it slowly} enough, a signal from the initial singularity has time to escape before enough mass gathers to trap it. Though the result above requires less regularity, we have stressed the globally $C^2$ case to make unequivocal the physical interpretation of the curvature and its energy condition. Note that it was also shown in \cite{wheelervaidya} that the naked singularity retains divergent curvature regardless of the form or smoothness of $m$.

The singularity at $(v,r) \to (0,0)$, then, is visible from infinity for an extended period provided that the {\it open}-appearing condition $\sup_v m'(v) < \frac{1}{16}$ is met. In a reasonable topology on the collection of these Vaidya spacetimes-- such as that induced by putting them in bijection with the set of admissible $m(v)$ endowed with the $C^1$ norm (note this topology has the minimal fineness in $m$ required to capture ``closeness" in $G$, and hence the implicit matter distribution, given (\ref{equation:curvature}))--, the subset exhibiting naked singularities has nonempty interior. While we have restricted $m$ to be required to level off at some $v_0$ for simplicity of presentation, this is not required for the most recent deduction (one may take a $C^1$ neighborhood around an $m(v)$ which does level off). In particular, these conclusions are not restricted to the asymptotically flat context. One hopes, of course, that this genericness is destroyed by perturbing outside of spherical symmetry. See \cite{wheelervaidya} for a more thorough and general investigation of these naked singularities.

\qed

\begin{figure}[t!]
\centering
\includegraphics[width=16cm]{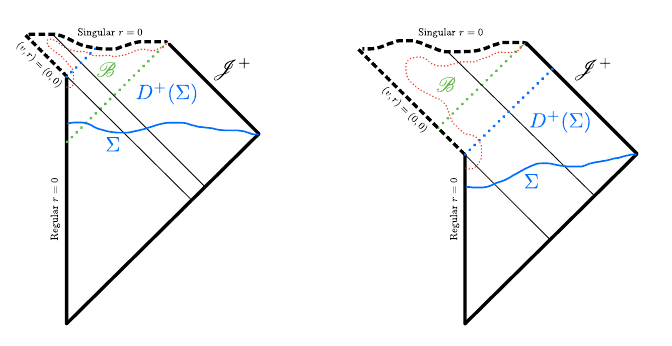}
\caption{\footnotesize The Penrose diagrams of the previous figure, with a complete space-like hypersurface $\Sigma$ added in each (in blue) to demonstrate the condition of Conjecture \ref{conjecture:globalcensor}. $D^+(\Sigma)$ is bounded by $\Sigma$ below and the blue dotted lines above. The red dotted curves mark the boundaries of typical singular neighborhoods. In the left case, the singular neighborhood shown has the property that its intersection with $D^+(\Sigma)$ is contained in $\mathscr{B}$; in the right case, any singular neighborhood intersects $D^+(\Sigma) \backslash \mathscr{B}$.}
\label{fig:vaidya2}
\end{figure}

This collection of Vaidya spacetimes is clearly in violation of the physical spirit of weak cosmic censorship, genericness aside. The standard IVP formulation, however, cannot consider them as such because they are outside of its scope due to their not utilizing a specific matter field characterized by a PDE coupled to the Einstein equation. That is: given a complete, asymptotically flat initial data set obtained from a complete space-like slice $\Sigma$ within one of these spacetimes, it is not clear what PDE one might use to evolve from it $D(\Sigma)$, so this evolution cannot be readily posed as an IVP. As they satisfy the dominant energy condition, however, one might argue that the Vaidya spacetimes have just as much a claim to exhibiting physical behavior as do toy matter models that happen to admit a PDE. In this sense, it is arguable that the IVP formulation of weak cosmic censorship, while certainly deeply significant, is not entirely sufficient to capture the physical spirit of the conjecture. We therefore put forward Conjecture \ref{conjecture:globalcensor} as an attempt at doing so.

It behooves us, then, to comment on how the naked singularities of Vaidya spacetimes indeed violate the condition of Conjecture \ref{conjecture:globalcensor}. A complete space-like hypersurface $\Sigma$ passes through the regular $r=0$ axis in the Minkowski region, say at some $v_\Sigma < 0$ (see Figure \ref{fig:vaidya2}). Each point $(v,0)$ satisfying $v_\Sigma < v < 0$ is in $D^+(\Sigma)$, but the entire regular $r=0, \; v<0$ axis is in the past of any of the escaping null geodesics found in Proposition \ref{proposition:nakedvaidya}, so these points are not in $\mathscr{B}$ (either heuristically in the spirit of Section \ref{section:blackhole} or within the formal structure of the abstract boundary, given Conjecture \ref{conjecture:blackcomplement}). Hence any singular neighborhood, which includes an open set around $(0,0)$ in the topology of the envelopment into $\mathbb{R} \times \mathbb{R}^3$ manifest in the coordinates (by Corollary \ref{corollary:envelopmentsingularneighborhood}, within the abstract boundary), must intersect $D^+(\Sigma) \backslash \mathscr{B}$, contrary to the condition in Conjecture \ref{conjecture:globalcensor}.

\section{Conclusions}
Through the definitions of Section \ref{section:blackhole}, we have provided a novel program for formalizing the notion of a black hole in an arbitrary maximal spacetime $(M,g)$, with no extrinsic structure or constricting hypotheses required, in a sense dual to the standard program of identifying that subset of $M$ which cannot be seen from infinity. We feel that this new program captures well important intuitive features a classical black hole should exhibit and nicely complements the existing treatments which are either much more limited in scope or hone in on other features of interest. Indeed, having provided one means of following through this program in a complete fashion, utilizing the tool of the abstract boundary to identify singular neighborhoods in Definition \ref{definition:singularneighborhood}, we have been able to arrive at several results pointing towards the naturalness of these definitions. 

As the abstract boundary is as unwieldy a tool as it is powerful, however, questions still remain. The largest such outstanding question regarding the physical reasonability of utilizing the abstract boundary formalism for the purpose of identifying black holes lies in Conjecture \ref{conjecture:separation}, by association with its immediate corollary in Conjecture \ref{conjecture:blackcomplement}. Other questions surrounding the relationship of the proposed notion of black holes to other, established intuitions include the link between $\mathscr{B}$ and trapped surfaces. While Penrose's Incompleteness Theorem ensures that trapped surfaces indicate null geodesic incompleteness, evidently the incompleteness need not be so severe as to indicate a nonempty $\mathscr{B}$ in a maximal extension (as seen in Kerr). Would some natural additional hypotheses, such as the spacetime's remaining globally hyperbolic upon maximally extending (i.e.\@ strong cosmic censorship), provide such an indication?

Beyond the philosophical appeal of providing a completely general characterization of the important physical concept of black holes unburdened by the constraint of asymptotic flatness, perhaps the most significant application of the program lies in its yielding a means to put forward a more comprehensive formulation of the Weak Cosmic Censorship Conjecture in Conjecture \ref{conjecture:globalcensor}. As seen in the explorations of Example \ref{example:vaidya}, this formulation is able to consider physically objectionable phenomena surrounding naked singularities that current IVP formulations cannot, both in and out of the asymptotically flat context. It is our hope that this will prove useful in rigorously illuminating the physical content of the General Theory of Relativity.

\section*{Acknowledgements}
The author would like to thank his advisor, Dr.\@ Hubert Bray, for supporting this work, as well as several others who have provided helpful commentary, discussion, and criticism throughout the development of this project, including Dr.\@ Marcus Khuri, Dr.\@ Demetre Kazaras, Dr.\@ Ronen Plesser, Dr.\@ Ben Whale, Dr.\@ Susan Scott, and Rui Xian Siew.

\section*{Statements and Declarations}
The authors have no relevant financial or non-financial interests to disclose. Data sharing is not applicable to this article as no datasets were generated or analysed during the current study.

\bibliographystyle{plain}
{\small \bibliography{references.bib}}

\begin{thebibliography}{10}

\bibitem{aghanim2020planck}
Nabila Aghanim, Yashar Akrami, Mark Ashdown, J~Aumont, C~Baccigalupi,
  M~Ballardini, AJ~Banday, RB~Barreiro, N~Bartolo, S~Basak, et~al.
\newblock Planck 2018 results-vi. cosmological parameters.
\newblock {\em Astronomy \& Astrophysics}, 641:A6, 2020.

\bibitem{ashley2002singularity}
Michael Ashley.
\newblock Singularity theorems and the abstract boundary construction.
\newblock 2002.

\bibitem{ashtekar2000generic}
Abhay Ashtekar, Christopher Beetle, Olaf Dreyer, Stephen Fairhurst, Badri
  Krishnan, Jerzy Lewandowski, and Jacek Wi{\'s}niewski.
\newblock Generic isolated horizons and their applications.
\newblock {\em Physical Review Letters}, 85(17):3564, 2000.

\bibitem{ashtekar2004isolated}
Abhay Ashtekar and Badri Krishnan.
\newblock Isolated and dynamical horizons and their applications.
\newblock {\em Living Reviews in Relativity}, 7(1):1--91, 2004.

\bibitem{bardeen1973four}
James~M Bardeen, Brandon Carter, and Stephen~W Hawking.
\newblock The four laws of black hole mechanics.
\newblock {\em Communications in mathematical physics}, 31(2):161--170, 1973.

\bibitem{barryscottsap}
Richard~A Barry and Susan~M Scott.
\newblock The strongly attached point topology of the abstract boundary for
  space-time.
\newblock {\em Classical and Quantum Gravity}, 31(12):125004, 2014.

\bibitem{bengtsson2011region}
Ingemar Bengtsson and Jose~MM Senovilla.
\newblock Region with trapped surfaces in spherical symmetry, its core, and
  their boundaries.
\newblock {\em Physical Review D}, 83(4):044012, 2011.

\bibitem{bernalsanchez}
Antonio~N Bernal and Miguel S{\'a}nchez.
\newblock On smooth cauchy hypersurfaces and geroch's splitting theorem.
\newblock {\em arXiv preprint gr-qc/0306108}, 2003.

\bibitem{bray2001proof}
Hubert~L Bray.
\newblock Proof of the riemannian penrose inequality using the positive mass
  theorem.
\newblock {\em Journal of Differential Geometry}, 59(2):177--267, 2001.

\bibitem{christodoulou1999instability}
Demetrios Christodoulou.
\newblock The instability of naked singularities in the gravitational collapse
  of a scalar field.
\newblock {\em Annals of Mathematics}, 149(1):183--217, 1999.

\bibitem{christodoulou1999global}
Demetrios Christodoulou.
\newblock On the global initial value problem and the issue of singularities.
\newblock {\em Classical and Quantum Gravity}, 16(12A):A23, 1999.

\bibitem{christodoulou1993global}
Demetrios Christodoulou and Sergiu Klainerman.
\newblock The global nonlinear stability of the minkowski space.
\newblock {\em S{\'e}minaire {\'E}quations aux d{\'e}riv{\'e}es partielles
  (Polytechnique) dit aussi" S{\'e}minaire Goulaouic-Schwartz"}, pages 1--29,
  1993.

\bibitem{curiel}
Erik Curiel.
\newblock The many definitions of a black hole.
\newblock {\em Nature Astronomy}, 3(1):27--34, 2019.

\bibitem{dafermos2014black}
Mihalis Dafermos.
\newblock Black holes without spacelike singularities.
\newblock {\em Communications in Mathematical Physics}, 332(2):729--757, 2014.

\bibitem{dafermos2013lectures}
Mihalis Dafermos and Igor Rodnianski.
\newblock Lectures on black holes and linear waves.
\newblock {\em Clay Math. Proc}, 17:97--205, 2013.

\bibitem{gerochboundary}
Robert Geroch.
\newblock Local characterization of singularities in general relativity.
\newblock {\em Journal of Mathematical Physics}, 9(3):450--465, 1968.

\bibitem{geroch1968singularity}
Robert Geroch.
\newblock What is a singularity in general relativity?
\newblock {\em Annals of Physics}, 48(3):526--540, 1968.

\bibitem{gerochgloballyhyperbolic}
Robert Geroch.
\newblock Domain of dependence.
\newblock {\em Journal of Mathematical Physics}, 11(2):437--449, 1970.

\bibitem{geroch1982singular}
Robert Geroch, Liang Can-bin, and Robert~M Wald.
\newblock Singular boundaries of space--times.
\newblock {\em Journal of Mathematical Physics}, 23(3):432--435, 1982.

\bibitem{harada2013threshold}
Tomohiro Harada, Chul-Moon Yoo, and Kazunori Kohri.
\newblock Threshold of primordial black hole formation.
\newblock {\em Physical Review D}, 88(8):084051, 2013.

\bibitem{hawking1965occurrence}
Stephen~W Hawking.
\newblock Occurrence of singularities in open universes.
\newblock {\em Physical Review Letters}, 15(17):689, 1965.

\bibitem{hawking1972black}
Stephen~W Hawking.
\newblock Black holes in general relativity.
\newblock {\em Communications in Mathematical Physics}, 25(2):152--166, 1972.

\bibitem{hellis}
Stephen~W Hawking and George Francis~Rayner Ellis.
\newblock {\em The large scale structure of space-time}, volume~1.
\newblock Cambridge university press, 1973.

\bibitem{hayward1994general}
Sean~A Hayward.
\newblock General laws of black-hole dynamics.
\newblock {\em Physical Review D}, 49(12):6467, 1994.

\bibitem{hayward2000black}
Sean~A Hayward.
\newblock Black holes: New horizons.
\newblock {\em arXiv preprint gr-qc/0008071}, 2000.

\bibitem{hayward2006formation}
Sean~A Hayward.
\newblock Formation and evaporation of nonsingular black holes.
\newblock {\em Physical review letters}, 96(3):031103, 2006.

\bibitem{huisken2001inverse}
Gerhard Huisken and Tom Ilmanen.
\newblock The inverse mean curvature flow and the riemannian penrose
  inequality.
\newblock {\em Journal of Differential Geometry}, 59(3):353--437, 2001.

\bibitem{joshi2000gravitational}
Pankaj~S Joshi.
\newblock Gravitational collapse: the story so far.
\newblock {\em Pramana}, 55(4):529--544, 2000.

\bibitem{joshi1992strong}
Pankaj~S Joshi and IH~Dwivedi.
\newblock Strong curvature naked singularities in non-self-similar
  gravitational collapse.
\newblock {\em General relativity and gravitation}, 24(2):129--137, 1992.

\bibitem{krolak1982definitions}
Andrzej Kr{\'o}lak.
\newblock Definitions of black holes without use of the boundary at infinity.
\newblock {\em General Relativity and Gravitation}, 14(8):793--801, 1982.

\bibitem{kuroda1984naked}
Yuhji Kuroda.
\newblock Naked singularities in the {V}aidya spacetime.
\newblock {\em Progress of theoretical physics}, 72(1):63--72, 1984.

\bibitem{lemos1992naked}
Jos{\'e}~PS Lemos.
\newblock Naked singularities: Gravitationally collapsing configurations of
  dust or radiation in spherical symmetry, a unified treatment.
\newblock {\em Physical review letters}, 68(10):1447, 1992.

\bibitem{luk2019strong}
Jonathan Luk and Sung-Jin Oh.
\newblock Strong cosmic censorship in spherical symmetry for two-ended
  asymptotically flat initial data ii: the exterior of the black hole region.
\newblock {\em Annals of PDE}, 5(1):1--194, 2019.

\bibitem{nielsen2009black}
Alex~B Nielsen.
\newblock Black holes and black hole thermodynamics without event horizons.
\newblock {\em General Relativity and Gravitation}, 41(7):1539--1584, 2009.

\bibitem{oneil}
Barrett O'neill.
\newblock {\em Semi-Riemannian geometry with applications to relativity}.
\newblock Academic press, 1983.

\bibitem{oppenheimer1939continued}
J~Robert Oppenheimer and Hartland Snyder.
\newblock On continued gravitational contraction.
\newblock {\em Physical Review}, 56(5):455, 1939.

\bibitem{penrosesingularity}
Roger Penrose.
\newblock Gravitational collapse and space-time singularities.
\newblock {\em Physical Review Letters}, 14(3):57, 1965.

\bibitem{penrose1969gravitational}
Roger Penrose.
\newblock Gravitational collapse: The role of general relativity.
\newblock {\em Nuovo Cimento Rivista Serie}, 1:252, 1969.

\bibitem{penrose1979singularities}
Roger Penrose.
\newblock Singularities and time-asymmetry.
\newblock In {\em General Relativity : An Einstein Centenary Survey},
  chapter~12, pages 581--683. Cambridge University Press, 1979.

\bibitem{penroseconformalinfinity}
Roger Penrose.
\newblock Republication of: Conformal treatment of infinity.
\newblock {\em General Relativity and Gravitation}, 43(3):901--922, 2011.

\bibitem{perlmutter1999measurements}
Saul Perlmutter, Goldhaber Aldering, Gerson Goldhaber, RA~Knop, Peter Nugent,
  Patricia~G Castro, Susana Deustua, Sebastien Fabbro, Ariel Goobar, Donald~E
  Groom, et~al.
\newblock Measurements of $\omega$ and $\lambda$ from 42 high-redshift
  supernovae.
\newblock {\em The Astrophysical Journal}, 517(2):565, 1999.

\bibitem{riess1998observational}
Adam~G Riess, Alexei~V Filippenko, Peter Challis, Alejandro Clocchiatti, Alan
  Diercks, Peter~M Garnavich, Ron~L Gilliland, Craig~J Hogan, Saurabh Jha,
  Robert~P Kirshner, et~al.
\newblock Observational evidence from supernovae for an accelerating universe
  and a cosmological constant.
\newblock {\em The Astronomical Journal}, 116(3):1009, 1998.

\bibitem{rodnianski2019naked}
Igor Rodnianski and Yakov Shlapentokh-Rothman.
\newblock Naked singularities for the {E}instein vacuum equations: The exterior
  solution.
\newblock {\em arXiv preprint arXiv:1912.08478}, 2019.

\bibitem{schmidt1991uniqueness}
Bernd Schmidt.
\newblock On the uniqueness of boundaries at infinity of asymptotically flat
  spacetimes.
\newblock {\em Classical and Quantum Gravity}, 8(8):1491, 1991.

\bibitem{schmidtbb}
BG~Schmidt.
\newblock A new definition of singular points in general relativity.
\newblock {\em General relativity and gravitation}, 1(3):269--280, 1971.

\bibitem{schoen1983existence}
Richard Schoen and S-T Yau.
\newblock The existence of a black hole due to condensation of matter.
\newblock {\em Communications in Mathematical Physics}, 90(4):575--579, 1983.

\bibitem{scottszekeres}
Susan~M Scott and Peter Szekeres.
\newblock The abstract boundary---a new approach to singularities of manifolds.
\newblock {\em Journal of Geometry and Physics}, 13(3):223--253, 1994.

\bibitem{scott2021endpoint}
Susan~M Scott and Ben~E Whale.
\newblock The endpoint theorem.
\newblock {\em Classical and Quantum Gravity}, 38(6):065012, 2021.

\bibitem{singh1996gravitational}
Tejinder~Pal Singh.
\newblock Gravitational collapse and cosmic censorship.
\newblock {\em arXiv preprint gr-qc/9606016}, 1996.

\bibitem{szekeres}
Gy{\"o}rgy Szekeres.
\newblock On the singularities of a riemannian manifold.
\newblock {\em Publicationes Mathematicae Debrecen 7}, 7:285, 1960.

\bibitem{wald1999gravitational}
Robert~M Wald.
\newblock Gravitational collapse and cosmic censorship.
\newblock In {\em Black holes, gravitational radiation and the universe}, pages
  69--86. Springer, 1999.

\bibitem{waldgr}
Robert~M Wald.
\newblock {\em General relativity}.
\newblock University of Chicago press, 2010.

\bibitem{whale2010dependence}
BE~Whale.
\newblock The dependence of the abstract boundary classification on a set of
  curves i: An algebra of sets on bounded parameter property satisfying sets of
  curves.
\newblock {\em arXiv preprint arXiv:1001.5091}, 2010.

\bibitem{whale2012dependence}
BE~Whale.
\newblock The dependence of the abstract boundary classification on a set of
  curves ii: How the classification changes when the bounded parameter property
  satisfying set of curves changes.
\newblock {\em arXiv preprint arXiv:1201.6414}, 2012.

\bibitem{whale2010foundations}
Ben~Edward Whale.
\newblock Foundations of and applications for the abstract boundary
  construction for space-time.
\newblock 2010.

\bibitem{wheelervaidya}
James Wheeler.
\newblock Generic naked singularities in vaidya spacetimes.
\newblock {\em Classical and Quantum Gravity}, 39(19):197001, September 2022.

\end{thebibliography}

\newpage

\appendix

\section{Review of the Abstract Boundary}

Here we provide a brief overview of the abstract boundary construction. We do this to establish ease of reference for lesser-known terms and notational standards used throughout this work, as well as to fill a gap in the literature on this subject which might obstruct its appreciation, namely a streamlined coverage of the abstract boundary in a case of particular interest, when $M$ is maximally extended, undistracted by various technicalities which are significant in general but not relevant to this case. We have also illustrated the definitions with the minimal example of $M = \mathbb{R}$, which is not worked out elsewhere, an immensely helpful case to see upon first encountering these concepts.

For a more detailed treatment, including a litany of examples addressing the various types of boundary points, we refer the reader to the seminal paper on the subject by Scott and Szekeres \cite{scottszekeres}, as well as the later paper by Barry and Scott introducing the a-boundary's strongly attached point topology \cite{barryscottsap}. We will follow the crucial features of those discussions closely, with some details and proofs omitted. Further resources on this topic include \cite{ashley2002singularity, whale2010dependence, whale2012dependence, whale2010foundations}, and references therein.

\subsection{The Essential Definitions}
The core construction of the abstract boundary may be carried out for any smooth manifold. The motivating idea that one should keep in mind is that the abstract boundary attempts to make rigorous and entirely chart-invariant the standard heuristic chart-based approach to identifying and analyzing singularities described at the beginning of sections \ref{section:blackhole} and \ref{section:singularnbhd}. All manifolds considered here will be smooth, connected, Hausdorff, paracompact, without boundary, and have the same dimension $n$ (with the case $n=4$ being of particular interest). The idea is to consider boundary points of $M$ under all possible smooth embeddings, quotienting by an equivalence relation capturing when boundary points arising in different embeddings are ``the same" from the perspective of $M$. To that end, we establish some standard notation. All definitions given in this and the following subsection were originally put forward in \cite{scottszekeres}.

\begin{definition}
An \ul{envelopment} is a smooth embedding $\phi: M \to \widehat M$, where $M$ and $\widehat{M}$ are smooth manifolds of the same dimension.
\end{definition}

As the dimensions of $M$ and $\widehat{M}$ are the same, $\phi(M)$ is an open submanifold of $\widehat{M}$. We will be interested in extracting information from the topological closure $\overline{\phi(M)} \subset \widehat{M}$ that is invariant under the choice of envelopment, keeping $M$ fixed.

\begin{definition}
Given an envelopment $\phi: M \to \widehat M$, a \ul{boundary set} $B$ is a non-empty subset of $\partial \phi(M) \subset \widehat{M}$.
\end{definition}

We must establish when such boundary sets are equivalent.

\begin{definition}\label{definition:cover}
Given boundary sets $B$, $B'$ under different envelopments $\phi: M \to \widehat{M}$, $\phi': M \to \widehat{M}'$, we say that $B$ \ul{covers} $B'$, written $B \triangleright B'$, if for every neighborhood $N \subset \widehat{M}$ of $B$, there exists a neighborhood $N' \subset \widehat{M}'$ of $B'$ such that 
$$ \phi \circ \phi'^{-1}(N' \cap \phi'(M)) \subset N.$$
\end{definition}

\begin{figure}[t]
\centering
\includegraphics[width=12cm]{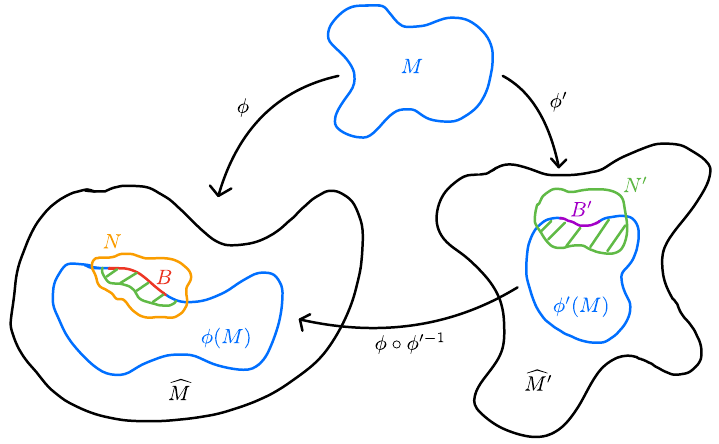}
\caption{ \footnotesize A depiction of the covering relation of Definition \ref{definition:cover}. Given any neighborhood $N \subset \widehat M$ of $B$, we may find a neighborhood $N' \subset \widehat M'$ of $B'$ ``contained" within $N$ under the mapping $ \phi \circ \phi'^{-1}$ between the envelopments.}
\label{fig:covering}
\end{figure}

The covering relation is depicted in Figure \ref{fig:covering}. This relation captures when $B$ contains all topological information present in $B'$, as it is shown in \cite{scottszekeres} (Theorem 19) that $B \triangleright B'$ iff every sequence $(p_n)_{n=1}^\infty \subset M$ with the property that $\phi'(p_n)$ has a limit point in $B'$ also satisfies that $\phi(p_n)$ has a limit point in $B$. We now say that boundary sets $B$ and $B'$ are {\it equivalent}, written $B \sim B'$, if both $B' \triangleright B$ and $B \triangleright B'$. It is not difficult to see that this establishes an equivalence relation on the collection of boundary sets, finally leading to:

\begin{definition}
An \ul{abstract boundary set} is an equivalence class of boundary sets, denoted $[B]$. An \ul{abstract boundary point} is an abstract boundary set which has a singleton $\{p\}$ as a representative boundary set in some envelopment. The \ul{abstract boundary} $\mathcal{B}(M)$ is the set of all abstract boundary points.
\end{definition}

A point of notation: we often refer to an abstract boundary point $[\{p\}]$ with representative boundary point $p \in \widehat M$ under some envelopment by the shorthand $[p]$.

It should be kept in mind that the primary utility of the abstract boundary does not lie in its capacity to provide at once an easily-conceivable global visualization of a spacetime together with its boundary (as some previous constructions sought to do)-- it is far too complicated an object for that. Instead, it provides a formalism for describing when boundary features are invariant under one's choice of envelopment, i.e. are intrinsic to $M$. Global visualizations of spacetimes are often achieved via envelopments (e.g.\@ in Penrose diagrams and global coordinate charts), and so it is of interest to know when features inferred from these visualizations are intrinsic to $M$. Clearly from the definitions, features which can be attributed to an abstract boundary set irrespective of its representative, as opposed to a particular boundary set, are manifestly independent of one's choice of envelopment. 

Before proceeding on to the classification, we illustrate the definitions with the single example for which a complete and nontrivial (if $M$ is compact, it admits no envelopments and hence has empty abstract boundary) description of the abstract boundary of a smooth manifold as a point-set is tractable to obtain.

\begin{figure}[t!]
\centering
\includegraphics[width=12cm]{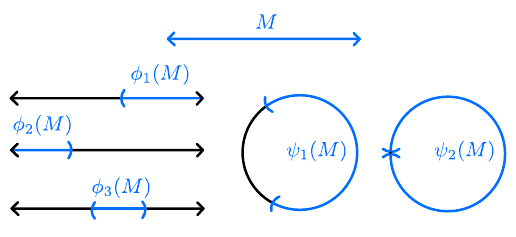}
\caption{\footnotesize The possible types of envelopments of $M = \mathbb{R}$ which give rise to boundary points. $M$ and its various images are displayed in blue, while the rest of the enveloping manifold $\widehat M$ is shown in black in each case. Types $\phi_1$ through $\phi_3$ and $\psi_1$ give rise to representatives of $[L]$ and $[R]$, while type $\psi_2$ gives rise to a representative of $[LR]$.}
\label{fig:Raboundary}
\end{figure}

\begin{example}\label{example:Raboundary}
The abstract boundary of $M = \mathbb{R}$. See Figure \ref{fig:Raboundary}.
\end{example}
We begin by considering that any envelopment sends $M$ either to an interval of the form $(a,b) \subset \mathbb{R}$, $-\infty \leq a < b \leq \infty$, or to a connected open strict subset of $S^1$, e.g.\@ $e^{i(c,d)} \subset S^1 \subset \mathbb{C}$, $0 < d-c \leq 2\pi$. Boundary sets in the former type of envelopment take the form $\{a\}, \{b\},$ or $\{a,b\}$ (any of these containing an $\infty$ would be omitted), while boundary sets in the latter type take the form $\{e^{ic}\}, \{e^{id}\}$, or $\{e^{ic}, e^{id}\}$ (in the special case $d-c = 2\pi$, these three are all the same). 

For two envelopments $\phi, \widetilde \phi : M \to \mathbb{R}$, the diffeomorphism $\phi \circ \widetilde \phi^{-1}: (\tilde a, \tilde b) \to {(a,b)}$ is either increasing or decreasing. It is straightforward to see that in the former case $\{a\} \sim \{\tilde a\}$ and $\{b\} \sim \{\tilde b\}$, while in the latter case $\{a\} \sim \{\tilde b\}$ and $\{b\} \sim \{\tilde a\}$, provided that the objects on both sides of the $\sim$ in each case are indeed boundary sets, i.e. are finite. Hence the envelopments into $\mathbb{R}$ give rise to exactly two abstract boundary points, which we will denote $[L]$ and $[R]$, indicating that they are the equivalence classes of the left and right endpoints of $\phi_0(M)$ under some particular chosen envelopment $\phi_0$ into the unit interval $(a_0,b_0) = (0,1)$.

It is similarly straightforward to observe that any envelopment $\psi : M \to S^1$ with the property that $d-c < 2\pi$ has that the pair $\{e^{ic}\}, \{e^{id}\}$ are equivalent to the pair $[L]$, $[R]$, so that these do not give any new abstract boundary points. However, when $d-c = 2\pi$, then $e^{ic} = e^{id}$, and in fact we now easily find that $\{e^{ic}\} \sim \{a_0,b_0\}$, so that $[\{a_0,b_0\}] = [\{e^{ic}\}]$ is also an abstract boundary point which we will denote by $[LR]$. Since we have exhausted all possible singleton boundary sets, the entirety of the abstract boundary of $M = \mathbb{R}$ is given by 
$$\mathcal{B}(\mathbb{R}) = \{[L],[R],[LR]\}.$$ 
\qed

One can appreciate that the next-simplest nontrivial case of $M = \mathbb{R}^2$ is already utterly impractical to describe completely by observing that $M$ is diffeomorphic to the unit disc $D^2$, and by identifying the points in the bounding $S^1$ according to a fundamental polygon of any closed surface $S$ whatsoever, one can obtain an evelopment of $M$ into $S$. All of these will yield different boundary topologies of $\partial \phi(M)$ and contain boundary points equivalent to various subsets of the original $S^1$ boundary, yielding a very large collection of distinct abstract boundary points. Even when working only with boundary points induced by envelopments of $M$ into $\mathbb{R}^2$, the Riemann mapping theorem ensures the possibilities are vast in that it guarantees the existence of such an envelopment with $\phi(M)$ being any non-empty simply connected open set in the plane.

\subsection{Classifying Abstract Boundary Points}

Thus far, the construction has used nothing more than the smooth manifold structure of $M$, and we have managed to generate the set $\mathcal{B}(M)$ of abstract boundary points. We would now like to use the spacetime structure of $(M,g)$ to classify the abstract boundary points according to a scheme which identifies any singularities. To avoid complications distracting from the physical application in mind, we will assume that $(M,g)$ is maximally extended subject to the desired regularity in $g$ and any other physical requirements (e.g.\@ time-orientability or other minimal causality condition). In the language of the broader classification detailed in \cite{scottszekeres}, this amounts to assuming there exist no {\it regular} boundary points. This considerably simplifies the forms of Definitions \ref{definition:infty}, \ref{definition:singularity}, and \ref{definition:mixedsingularity} below as compared to their statements in \cite{scottszekeres}.

As is standard in the theory of singularities, we must distinguish singular boundary points from those at ``infinity" or otherwise by identifying singularities as being reachable with bounded parameter via a physically meaningful family of curves. The minimum requirements of the curve family are as follows:

\begin{definition} \label{definition:bpp}
A family $\mathcal{C}$ of $C^1$ parameterised curves $\gamma: [a,b) \to M$ ($a < b \leq \infty$) is said to have the \ul{bounded parameter property} (b.p.p.) provided that
\begin{enumerate}[(i)]
\item For any $p \in M$, there exists a $\gamma \in \mathcal{C}$ passing through $p$.
\item For any $\gamma \in \mathcal{C}$, any subcurve $\gamma|_{[a',b')}$ (where $ [a',b') \subset [a,b)$) is also in $\mathcal{C}$.
\item For any $\gamma, \tilde \gamma \in \mathcal{C}$ related by an increasing change in parameter, either both $b, \tilde{b} < \infty$ ($\gamma, \tilde \gamma$ have \ul{bounded parameter} ) or $b = \tilde{b} = \infty$ ($\gamma, \tilde \gamma$ have \ul{unbounded parameter} ). 
\end{enumerate}
\end{definition}

These conditions simply ensure that the family $\mathcal{C}$ has sufficiently many curves to arguably characterize boundary behavior, and that it provides unambiguous notions of ``bounded" and ``unbounded" distance along the curves. Families of particular interest that satisfy this definition include the collection of geodesics with affine parameter, the collection of $C^1$ curves with generalized affine parameter, the causal or future-directed causal subfamilies of each of these, and the collection of $C^1$ timelike curves with proper time parameter. The  b.p.p.\@ family of curves with which one chooses to work is the means by which the spacetime metric characterizes the upcoming classification-- notice that in all of the curve families just listed, the metric determines the parameterization, and hence whether a given curve has bounded or unbounded parameter.

\begin{definition}
Given an envelopment $\phi: M \to \widehat{M}$ and a b.p.p.\@ family of curves $\mathcal{C}$, a boundary set $B \subset \partial \phi(M)$ is \ul{$\mathcal{C}$-approachable}, or just \ul{approachable}, if there exists a $\gamma \in \mathcal{C}$ such that $\phi \circ \gamma$ has a limit point in $B$.
\end{definition}

It is a direct corollary of the sequential remark made following Definition \ref{definition:cover} that if $B \triangleright B'$ and $B'$ is approachable, then $B$ is approachable as well (via the same curve $\gamma$). Hence the property of approachability passes to the abstract boundary: if $B \sim B'$, then $B$ is approachable iff $B'$ is approachable, and it is unambiguous to say that the abstract boundary set $[B]$ is approachable. Approachability is a basic requirement of being able to proceed with the classification, as approaching curves are the means by which boundary points are measured to be at ``finite" or ``infinite" distance, and this is precisely what determines whether the boundary point is deemed singular or not:

\begin{definition}\label{definition:infty}
Given an envelopment $\phi: M \to \widehat{M}$ ($M$ maximally extended) and a b.p.p.\@ family of curves $\mathcal{C}$, an approachable boundary point $p \in \widehat{M}$ will be called a \ul{point at infinity} if every $\gamma \in \mathcal{C}$ approaching $p$ has unbounded parameter.
\end{definition}

\begin{definition}\label{definition:singularity}
Given an envelopment $\phi: M \to \widehat{M}$ ($M$ maximally extended) and a b.p.p.\@ family of curves $\mathcal{C}$, an approachable boundary point $p \in \widehat{M}$ will be called \ul{singular} or  a \ul{singularity} if there exists a $\gamma \in \mathcal{C}$ approaching $p$ with bounded parameter.
\end{definition}

Again, the properties of being a point at infinity or a singularity clearly pass to the abstract boundary, so that we may say the abstract boundary point $[p]$ is a point at infinity or singularity in the same manner. This nominally achieves the desired classification, but the immense flexibility of envelopments means that we need one final distinction to extract ``pure" singular behavior:

\begin{definition}\label{definition:mixedsingularity}
A singular boundary point will be called \ul{mixed} or \ul{directional} if it covers a point at infinity. Otherwise, it will be called a \ul{pure singularity}.
\end{definition}

\begin{figure}[b!]
\centering
\includegraphics[width=10cm]{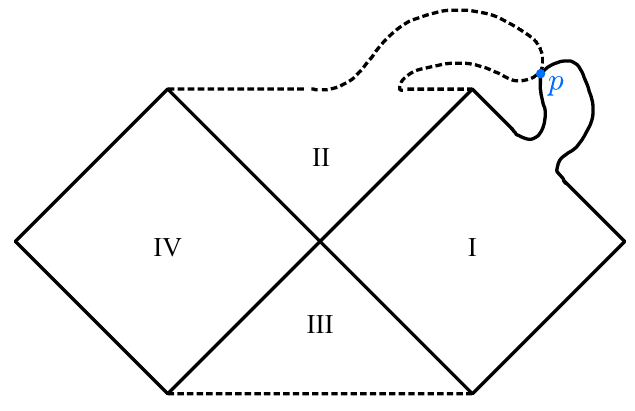}
\caption{\footnotesize An envelopment of the Schwarzschild spacetime giving rise to a mixed singularity. The usual envelopment of the Penrose diagram is locally modified around a portion of the $r=0$ boundary to region II and a portion of $\mathscr{J}^+$ for region I, stretching the boundaries until they touch at a point. The Schwarzschild spacetime can still be smoothly (not conformally) embedded without issue into the interior of this region, and the boundary point $p$ of tangency is a mixed singularity. Thus if Definition \ref{definition:singularneighborhood} were based around general singularities instead of pure singularities, then singular neighborhoods would apparently have to surround $\mathscr{J}^+$ as well, which is undesirable.}
\label{fig:mixedsingularity}
\end{figure}

Mixed singularities are unaviodable byproducts of the generality of the abstract boundary construction. They apparently contain both singular and nonsingular behavior, and we'd like to filter out the latter in our identification of singular neighborhoods. See Figure \ref{fig:mixedsingularity} for an example with the Schwarzschild spacetime demonstrating the need for this distinction, particularly as it impacts the notion of singular neighborhood given in Definition \ref{definition:singularneighborhood}. This definition completes the classification scheme of abstract boundary points for a maximally extended spacetime. Within it, every abstract boundary point is exactly one of the following, given a choice of b.p.p.\@ curve family: unapproachable, a point at infinity, a mixed singularity, or a pure singularity.

It is worth briefly commenting on how one might hope to identify a pure singularity $p \in \widehat M$, as it in principle requires checking the status of boundary points in every other envelopment and their covering relations with $p$. The easiest way to identify a pure singularity via the data available in a single envelopment is to observe that if {\it every} $\gamma \in \mathcal{C}$ approaching $p$ has bounded parameter, then $p$ must be a pure singularity. This is simply because if $p$ covered a point at infinity $q$, then any curve $\gamma$ approaching $q$ would also have to approach $p$, and this curve would have to have unbounded parameter. This is the easiest way to identify, say, the $r=0$ points in a Schwarzschild Penrose diagram as necessarily being pure singularities with respect to physically relevant curve families.

\subsection{The Strongly Attached Point Topology}

We now turn to the problem of identifying a natural topology on the set $\overline{M} := M \cup \mathcal{B}(M)$, which is furnished readily by the available topological structure of the envelopments giving rise to abstract boundary sets. The definitions in this section were originally put forward in \cite{barryscottsap}. We first define two terms capturing what it means for a boundary set to be ``near" an open set in $M$-- the former of these we state only to simplify the statement of some results in Section \ref{section:results}, while the latter is critical to identifying singular neighborhoods.

\begin{definition}\label{definition:attached}
Given an open set $U \subset M$ and an envelopment $\phi : M \to \widehat{M}$, a boundary set $B \subset \partial \phi(M)$ is said to be \ul{attached} to $U$ if $B \cap \overline{\phi(U)} \neq \emptyset$.
\end{definition}

\begin{definition}\label{definition:stronglyattached}
Given an open set $U \subset M$ and an envelopment $\phi : M \to \widehat{M}$, a boundary set $B \subset \partial \phi(M)$ is said to be \ul{strongly attached} to $U$ if there exists an open set $N \subset \widehat{M}$ containing $B$ such that
$$N \cap \phi(M) \subset \phi(U).$$
\end{definition}

\begin{figure}[t]
\centering
\includegraphics[width=14cm]{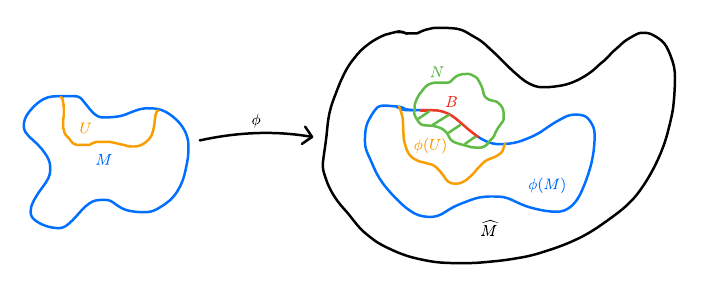}
\caption{ \footnotesize A depiction of Definition \ref{definition:stronglyattached}. We may find an open neighborhood $N \subset \widehat M$ around the boundary set $B \subset \partial \phi(M) \subset \widehat M$ which has intersection with $\phi(M)$ entirely contained within $\phi(U)$. $B$, and every boundary point therein, is strongly attached to $U$.}
\label{fig:stronglyattached}
\end{figure}

See Figure \ref{fig:stronglyattached}. This latter definition captures that $U$ ``surrounds" $B$ as much as possible from within $M$, in the sense that if a sequence $(p_n)_{n=1}^\infty \subset M$ has the property that $(\phi(p_n))$ limits to a point in $B$, then $(p_n)$ eventually enters and remains inside $U$. What's more, the collection of open sets in $M$ to which $B$ is strongly attached fully characterizes the topological association of $B$ to $M$, as seen in the near-converse to the above: if $(p_n)$ eventually enters and remains inside every open set $U \subset M$ to which $B$ is strongly attached, then $(\phi(p_n))$ has a limit point in $B$.

It is a fairly immediate exercise in definition chasing to verify that if the boundary set $B$ is strongly attached to $U \subset M$ and $B \triangleright B'$, then $B'$ is also strongly attached to $U$. Hence the property of being strongly attached to a given open set passes to the abstract boundary, and it is unambiguous to say that the abstract boundary set $[B]$ is strongly attached to $U$. Denoting by $\mathcal{B}_U \subset \mathcal{B}(M)$ the set of abstract boundary points strongly attached to $U$, we may now characterize the desired topology on $\overline{M}$:

\begin{definition}\label{definition:stronglyattachedtopology}
The \ul{strongly attached point topology} $\; \mathcal{T}_{sap}(M)$ on $\overline{M} = M \cup \mathcal{B}(M)$ is the topology with basis
$$\mathcal{W} := \{U \cup \mathcal{B}_U \; | \; U \subset M \text{ is open} \}.$$
\end{definition}

It is straightforward to see that $\mathcal{W}$ covers $\overline{M}$ and that $ \mathcal B_{U_1} \cap \mathcal B_{U_2} = \mathcal B_{U_1 \cap U_2}$ (\cite{barryscottsap}, Lemmas 17 and 22), so that $\mathcal{W}$ is indeed a legitimate basis to a topology. This topology inherits from the sequential observations made following Definition \ref{definition:stronglyattached} the property that, given a boundary point $p \in \partial{\phi(M)}$ under the envelopment $\phi: M \to \widehat M$, a sequence $(p_n)_{n=1}^\infty \subset M \subset \overline{M}$ limits to the abstract boundary point $[p] \in \overline{M}$ iff $(\phi(p_n))$ limits to $p \in \widehat{M}$. As it can be shown that $\mathcal{T}_{sap}(M)$ is first countable (\cite{barryscottsap}, Proposition 46), and hence sequential, this indicates the strong association between the topological structure of $\mathcal{T}_{sap}(M)$ and that of the various envelopments.

Additional features of interest of $\mathcal{T}_{sap}(M)$ are that it induces the standard topology back onto $M$, $M \subset \overline{M}$ is open, and the closure of $M$ is indeed all of $\overline{M}$. Hence $\mathcal{B}(M)$ is the topological boundary $\partial M$ of $M$ in $\overline{M}$ under $\mathcal{T}_{sap}(M)$, as one would hope of a reasonable topology on $\overline{M}$. While $\overline{M}$ is {\it not} Hausdorff, this is a feature rather than a bug of $\mathcal{T}_{sap}(M)$: the failure of Hausdorff separation between distinct abstract boundary points encodes when boundary points in different envelopments are topologically entangled, e.g.\@ when one covers the other but not vice versa, or when a sequence in $M$ may limit to both simultaneously under their respective envelopments. As seen in the discussion following Example \ref{example:Raboundary} above, there is immense degeneracy in distinct abstract boundary points covering others, and such points' topological relations to $M$ are intertwined in a complicated manner, so it is useful and expected that $\mathcal{T}_{sap}$ should encode this in topological closeness. Also see Example \ref{example:Rsap} below. The important separation property for our purposes is that points in $M$ {\it are} Hausdorff separated from points in $\mathcal{B}(M)$. For proof and additional discussion of these features and more, we again refer the reader to \cite{barryscottsap}.

While we generally refer to \cite{barryscottsap} for proofs of a-boundary topological features, we will now provide one example of how such arguments go by extending a result in \cite{barryscottsap} indicating the close topological relationship between $\mathcal{T}_{sap}(M)$ and envelopments. Given an envelopment $\phi: M \to \widehat{M}$, we denote by $\sigma_\phi \subset \mathcal{B}(M)$ the set of abstract boundary points with singleton representatives in $\partial \phi(M)$, e.g.
$$\sigma_\phi := \{[p] \in \mathcal{B}(M) \; | \; p \in \partial \phi(M)\}. $$
Since distinct boundary points in the same envelopment cannot cover each other due to the Hausdorff property of $\widehat{M}$, clearly $\sigma_\phi$ is in bijection with $\partial \phi(M)$ under the natural quotient map $\pi : \partial \phi(M) \to \sigma_\phi$ given by $\pi(p) = [p]$.

\begin{proposition}
When $\sigma_\phi \subset \mathcal{B}(M)$ and $\partial \phi(M)$ are given the subspace topologies induced by $\mathcal{T}_{sap}(M)$ and $\widehat M$ respectively, the quotient map $\pi: \partial \phi(M) \to \sigma_\phi$ is a homeomorphism.
\end{proposition}

\proof
Let $O \subset \sigma_\phi$ be open in the subspace topology, so that $O = \widetilde U \cap \sigma_\phi$ for some open $\widetilde U \subset \overline{M}$. WLOG, we may assume $\widetilde U$ is in the basis, i.e.\@ $\widetilde U = U \cup \mathcal{B}_U$ for some open $U \subset M$, and hence that $O = \mathcal{B}_U \cap \sigma_\phi$. Take $p \in \pi^{-1}(O)$, so that $p \in \partial \phi(M)$ such that $[p] \in O \subset \mathcal{B}_U$, and hence $p$ is strongly attached to $U$. Then there exists a neighborhood $N \subset \widehat{M}$ of $p$ such that $N \cap \phi(M) \subset \phi(U)$. Of course, every point $q \in N \cap \partial \phi(M)$ is also strongly attached to $U$, as $N$ serves equally well as the required neighborhood of $q$ in Definition \ref{definition:stronglyattached}. That is, $\pi(N \cap \partial \phi(M)) \subset \mathcal{B}_U \cap \sigma_\phi = O$, and hence $N \cap \partial \phi(M)$ is a neighborhood of $p$ entirely contained in $\pi^{-1}(O)$, showing $\pi^{-1}(O)$ is open. This shows $\pi$ is continuous.

Now let $O \subset \partial \phi(M)$ be open in the subspace topology, so that $O = N \cap \partial \phi(M)$ for some open $N \subset \widehat{M}$. Fix $p \in O$, and observe that since $\widehat M$ is regular, there exists an open neighborhood $N'$ of $p$ with $\overline{N'} \subset N$. Setting $U := \phi^{-1}(N' \cap \phi(M))$ and noting $p$ is clearly strongly attached to $U$, we have that $\mathcal{B}_U \cap \sigma_\phi = \left( U \cup \mathcal{B}_U \right) \cap \sigma_\phi$ is an open neighborhood of $[p] = \pi(p)$ in $\sigma_\phi$. Noting that $q \in \partial \phi(M)$ being strongly attached to $U$ implies that $q \in \overline{\phi(U)}$, we have that 
$$\pi^{-1}(\mathcal{B}_U \cap \sigma_\phi) \subset \overline{\phi(U)} \cap \partial \phi(M) \subset \overline{N'} \cap \partial \phi(M) \subset N \cap \partial \phi(M) = O.$$
We therefore have $\mathcal{B}_U \cap \sigma_\phi \subset \pi(O)$, so that $\mathcal{B}_U \cap \sigma_\phi$ is a neighborhood of $\pi(p)$ entirely contained in $\pi(O)$, showing $\pi(O)$ is open and hence that $\pi$ is an open map.

\qed

This strengthens Proposition 52 in \cite{barryscottsap}, which invoked an additional, somewhat complicated hypothesis (given there as Condition 51) to prove an equivalent result. Extracting the intuitive information from this result requires a slight rephrasing. Given an envelopment $\phi: M \to \widehat{M}$, define $\overline \pi : \overline{\phi(M)} \to \overline{M}$ by

$$ \overline{\pi}(p) =
 \begin{cases} 
 	\phi^{-1}(p) & p \in \phi(M) \\
      [p] & p \in \partial \phi(M).
\end{cases} $$
This map just identifies points in $\overline{\phi(M)} \subset \widehat{M}$ with points in $\overline{M}$ in the most direct way possible. Essentially the same proof as above carries through to demonstrate the following:

\begin{proposition}\label{proposition:envelopmentembedding}
When $\overline{\phi(M)}$ is given the subspace topology induced by $\widehat M$, the map $\overline \pi:  \overline{\phi(M)} \to \overline{M}$ is a topological embedding, i.e. a homeomorphism onto its image.
\end{proposition} 
\qed

This essentially says that the natural identification of the closure of $M$ within an envelopment with the corresponding points in $\overline{M}$ preserves all topological features of the envelopment. The abstract boundary endowed with the strongly attached point topology, then, naturally contains within it all possible closures of $M$ within envelopments, complete with their topological structure. 

We once again conclude by demonstrating the definitions via the single example for which complete, explicit, and nontrivial computation is tractable.

\begin{example}\label{example:Rsap}
The strongly attached point topology on $\mathcal{B}(M)$ for $M= \mathbb{R}$.
\end{example}
The subspace topology $\mathcal{T}_{sap}^{\mathcal{B}}(\mathbb{R})$ on $\mathcal{B}(\mathbb{R})$ induced by $\mathcal{T}_{sap}(\mathbb{R})$ on $\overline{\mathbb{R}}$ has basis given by

$$\mathcal{W}^\mathcal{B} := \{ \mathcal{B}_U \; | \; U \subset \mathbb{R} \text{ is open}\}.$$
It is straightforward to see that if an open set $U \subset \mathbb{R}$ contains an interval of the form $(a,\infty)$, then $[R]$ is strongly attached to $U$; if it contains an interval of the form $(-\infty,b)$, then then $[L]$ is strongly attached to $U$; and if it contains intervals of both forms, then all of $[L],[R],$ and $[LR]$ are strongly attached to $U$. Hence we have

$$\mathcal{W}^\mathcal{B} =  \{ \; \{[L]\}, \{[R]\}, \{[L],[R],[LR]\} \; \},$$
and this yields

$$\mathcal{T}_{sap}^{\mathcal{B}}(\mathbb{R}) =  \{ \; \emptyset, \{[L]\}, \{[R]\}, \{[L],[R]\}, \{[L],[R],[LR]\} \; \}.$$
This topology is $T_0$ but not $T_1$. That every neighborhood of $[LR]$ includes both $[L]$ and $[R]$ encodes topologically that $[LR] \triangleright [L]$ and $[LR] \triangleright [R]$.

\end{document}